\documentstyle[epsfig,12pt]{article}

\begin{document}

\title{A Muon Collider Scheme Based on Frictional Cooling }
\date{}
\author{H. Abramowicz~\dag, A. Caldwell~\ddag, R. Galea~\S,  S. Schlenstedt~\P, \\
\small
\it{\dag\ Tel Aviv University, Tel Aviv, Israel.}\\
\small
\it{\ddag\ Max-Planck-Institut f\"{u}r Physik, Munich, Germany.}\\
\small
\it{\S\ Nevis Laboratories, Columbia University, Irvington, NY, USA.}\\ 
\small
\it{\P\  DESY, Zeuthen, Germany.}}
\maketitle

\begin{abstract}

Muon Colliders would usher in a new era of
scientific investigation in the field of high energy particle physics.
The cooling of muon beams is proving to be
the greatest obstacle in the realization of a Muon Collider. Monte
Carlo simulations of a muon cooling scheme based on 
Frictional Cooling were performed. Critical issues, which
require further 
study, relating to the
technical feasibility of such a scheme are identified.
Frictional Cooling, as outlined in this paper,
provides sufficient six-dimensional emittance  to make luminous
collisions possible. It holds exciting potential in solving the
problem of Muon Cooling.

\end{abstract}

\section{Introduction}
\subsection{Motivation for a Muon Collider}
\label{sec:motivation}

The principal motivation for building a Muon Collider is to extend the
high energy frontier of lepton colliders.  With LEP, we likely have
reached  
the maximum energy achievable with a circular electron collider, i.e., about
$100$~GeV/beam.  The limitation is the energy lost each turn through
synchrotron radiation, which scales as $(E/m)^4$, where $E$ is the
energy of the beam and $m$ is the mass of the particle.  
A linear electron-positron collider is currently under
consideration which will operate at energies of $\le 500$~GeV per
beam.  Reaching 
these energies requires very high electric fields as well as
very long accelerators.  Barring a breakthrough in acceleration techniques, it
is difficult to see how much higher energies will be possible.

A Muon Collider could reach much higher energies in a 
circular accelerator since $m_{\mu}\approx 206~ m_{e}$.  Collisions between
point-like particles would then be possible at energies similar to those
achievable for proton accelerators, thus reopening the energy frontier
with lepton  
colliders.  The physics case for a high energy Muon
Collider has been reviewed in many places.  A recent compilation can be
found in~\cite{ref:colliderstudy}.

A significant additional benefit of a Muon Collider would be the
intense and well understood 
flux of neutrinos resulting from muon decays allowing long baseline
neutrino oscillation experiments to be performed with high statistics.
The Collider complex 
could also 
be designed in such a way that neutrino scattering physics programs could be
pursued.  The physics case for a Neutrino Factory has
been studied extensively and detailed accelerator studies are
underway~\cite{ref:STUDYII}. 
Low energy muons have a wide range of physics 
applications~\cite{ref:lowenergymu} which could also be pursued if the
intense muon source required for the Muon Collider were constructed.

\subsection{Difficulties}

An obvious difficulty for the construction of a Muon Collider is the
short muon lifetime ($2.2$~$\mu$s).  Muons are not readily available and
must be produced with a multi-MW proton source.  Once the muons are produced,
they occupy a very large phase space which must be reduced.  The time
available for this process is too short for existing techniques to be
applicable (i.e.; stochastic cooling, synchrotron radiation damping
rings).  A new 
technique 
is required to reduce the muon emittance in a very short time. A
related problem is that of backgrounds resulting from muon decays.
Muon decays 
at all stages of the accelerator complex will produce radiation
levels which will pose significant challenges for the accelerator
elements.  Any detector used to study the muon collisions will also
see a 
large background flux from upstream muon decays.
Of particular importance for a high energy, high current, muon beam
is the radiation hazard resulting from the subsequent neutrino beam, which 
cannot be shielded~\cite{ref:neutrinoradiation}.  

\subsection{Muon Cooling}

Many of the difficulties listed above could be reduced with  very effective
muon cooling.  Cooler beams would allow fewer muons for a given
luminosity, and thereby reduce the experimental backgrounds, reduce the
radiation from muon decays, reduce the radiation from neutrino interactions,
and allow for smaller apertures in the accelerator elements. Fewer
required muons would  also be welcome for the targetry and proton
driver, which provide severe technical and financial constraints. We
therefore 
identify the efficient cooling of the muon beam phase space as the
signature challenge for a future Muon Collider.

\section{Cooling Schemes}
\subsection{Overview of a Muon Collider Complex}
\label{sec:overview}

\begin{figure}
\begin{center}
\epsfig{file=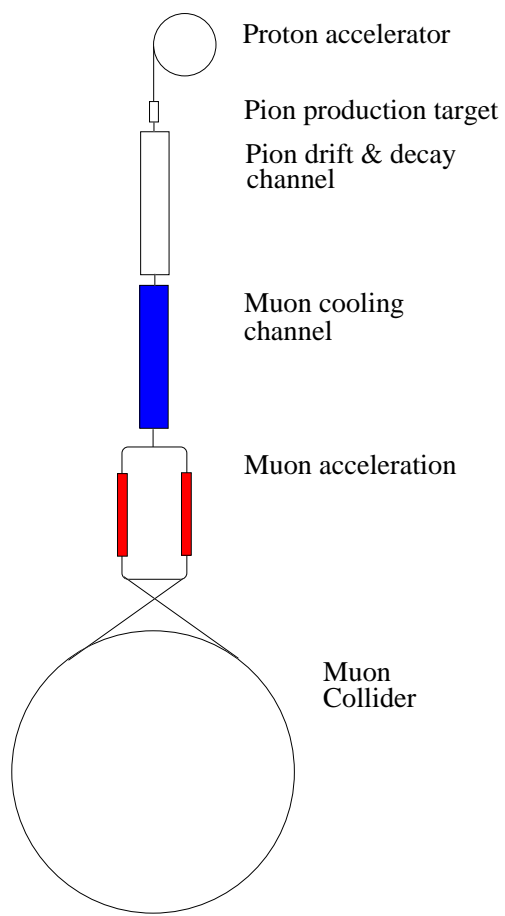,height=12.cm}
\end{center}
\caption[]{Schematic of a Muon Collider complex.}
\label{fig:Overview}
\end{figure}

Before briefly reviewing the proposed muon cooling schemes, we provide an
overview of a possible Muon Collider complex along with beam parameters.
Figure~\ref{fig:Overview} shows schematically the different elements 
required for a Muon Collider.  
The first stage is a proton driver, which is
a high intensity proton accelerator producing several MW of protons in
the energy range $2-30$~GeV (different schemes have different optimal 
energies).  The protons then impact a target in a region with a very strong
magnetic field, and the resulting charged particles, primarily pions,
are captured in a decay 
channel.  The pion drift and decay channel is optimized  to yield a
maximum number of muons in the required energy range.  The muons then
enter a cooling channel where the phase space of the beam is
compressed and they are then brought to the desired energy in a series
of 
accelerators.

A parameter
set for a 3~TeV center-of-mass (CoM) energy
Muon Collider is given in~\cite{ref:Ankenbrandt}, and is
summarized in Table~\ref{tab:pars}.
\begin{table}[hbpt]
\begin{center}
\begin{tabular}{lc}
\hline 
CoM energy (TeV) & 3 \\
proton energy (GeV) & 16 \\
protons/bunch & $2.5\cdot 10^{13}$ \\
bunches/fill & 4 \\
Rep. rate (Hz)  & 15 \\
$\mu$/bunch & $2 \cdot 10^{12}$ \\
collider circ. (m) & 6000 \\
$\epsilon_{6,N} (\pi {\rm m})^3$ & $ 1.7 \cdot 10^{-10} $ \\
Luminosity (cm$^{-2}$s$^{-1}$) & $7 \cdot 10^{34}$  \\
\hline
\end{tabular}
\caption{A set of parameters for a 3 TeV CoM energy Muon 
Collider~\cite{ref:Ankenbrandt}.}
\label{tab:pars}
\end{center}
\end{table}
The required normalized six-dimensional emittance for this  parameter set is 
$\epsilon_{6,N}=1.7\cdot10^{-10}$ in units of $({\pi {\rm m}})^3$.  
At the end of the pion drift and decay channel, the RMS dimensions of
the muon cloud  in spatial coordinates 
x, y, z are approximately 0.05, 0.05, 10~m, and in momentum coordinates
$P_x, P_y, P_z$ are   
$50, 50, 100$~MeV/c, yielding $\epsilon_{6,D}=1.7\cdot 10^{-4} ({\pi
{\rm m}})^3$.
The required emittance reduction is therefore of order $10^6$, and
this reduction 
must be accomplished with a reasonable efficiency. 

To reach the luminosities given in Table~\ref{tab:pars}, $0.08~\mu$
per $16$~GeV 
proton must 
be in the cooled phase space for each muon charge.   If we fix the
power of the proton driver at 
$4$~MW, then we aim for $0.01$ muons of each sign per $2$~GeV
proton, which is our optimal proton beam energy (see section~\ref{sec:yields}). 

The cooling scheme which has received considerable  consideration is
Ionization 
Cooling.  We describe this briefly in the next section.  The cooling scheme
we have chosen to investigate, Frictional Cooling, is then broadly
outlined.  The remainder of this paper describes in more detail a
conceptual design for the early stages of a Muon Collider based on
Frictional Cooling. It should be understood as an attempt to find a
solution to the cooling problem 
satisfying physical limitations such as muon lifetime, scattering
effects and energy loss mechanisms. Technical feasibility has not been
investigated. 

\subsection{Ionization Cooling}

Ionization Cooling operates at kinetic energies of the order of
100~MeV. For a Muon
Collider, emittance
reduction would be  achieved in alternating steps.
First, transverse emittance cooling is achieved by lowering the muon beam
energy in high density absorbers (such as liquid hydrogen) and then
reaccelerating the beam to keep the average energy constant.
Both transverse and longitudinal momenta are reduced in the absorbers,
but only the longitudinal momentum is restored by the RF. In this way,
transverse cooling can be achieved up to the limit given by multiple
scattering. Second,
longitudinal emittance cooling is achieved by adding dispersion into the muon
beam and passing the momentum gradiented beam through wedge shaped
absorbers such that the higher momentum muons pass through more
absorber than the lower momentum muons. This process adds to the
transverse emittance. The steps can then be repeated until the limits
from multiple scattering are reached. The trade off between the
transverse and longitudinal cooling portions is called {\it emittance
exchange} and the resultant beam is cooled in six
dimensions. 

Although Ionization Cooling shows promise and has been
investigated in detail~\cite{ref:ionization}, the simulation studies
have yet to show 
the required emittance 
reduction to make a Muon Collider feasible. Various schemes based on
Ionization Cooling have achieved six-dimensional emittance reductions
of the order of 100-200~\cite{ref:nufact02}.

\section{Frictional Cooling}
\subsection{Basic Idea}
\begin{figure}
\begin{center}
\epsfig{file=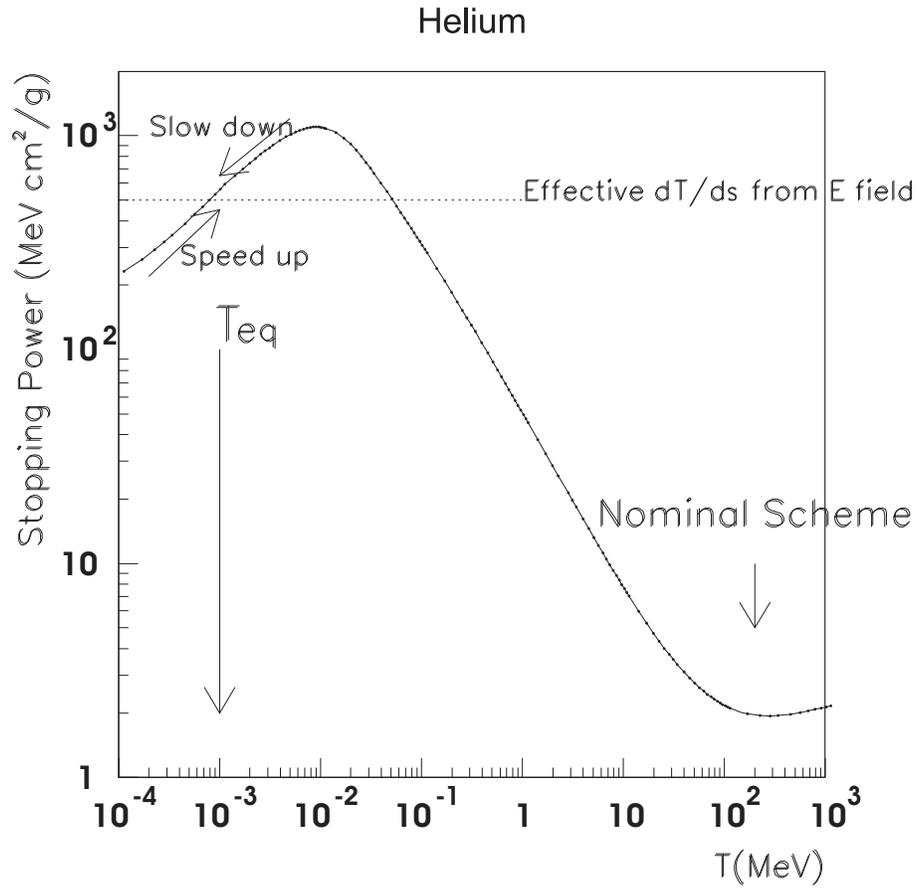,width=12.cm}
\end{center}
\caption[]{The stopping power, $\frac{1}{\rho}dT/ds$, in Helium as a
function of kinetic energy, $T$, 
for $\mu^+$ (solid line).  The effective stopping power resulting from
an external 
electric field is sketched as a dotted line.  An equilibrium kinetic
energy, $T_{eq}$, near $1$~keV would result. The nominal scheme discussed for a
Neutrino Factory, based on Ionization Cooling, would cool muons near
$T=200$~MeV.} 
\label{fig:dEdx}
\end{figure}
The basic idea of Frictional Cooling~\cite{ref:PSI} 
is to bring the muons into a
kinetic energy range, $T$, where energy loss per unit path length, $ds$,
increases with 
kinetic energy.  A constant accelerating
force is then applied to the muons resulting
in an equilibrium kinetic energy.  A sample $\frac{1}{\rho} dT/ds$
curve is shown in 
Fig.~\ref{fig:dEdx}, where it is seen that this condition can be met for
kinetic energies below a few keV, or kinetic energies beyond about $200$~MeV.
At the high energy end, the change in $dT/ds$ with energy is only logarithmic,
whereas it is approximately proportional to speed at low energies. We
focus on the region below the ionization peak. In this region,
ionization is suppressed because the relative speed between the muon
and orbital electrons is too low to allow sufficient transfer of
kinetic energy for ionization.

\subsection{Some Difficulties \& Solutions}
We study the low energy regime where $dT/ds$ is proportional to the
speed $v$ and $v \leq \alpha c$,
where $\alpha$ is the fine structure constant.
In this energy regime, we imagine applying an electric field which 
compensates $dT/ds$, yielding an equilibrium kinetic energy.
Several issues are immediately apparent:

\begin{itemize}
\item $\frac{1}{\rho}dT/ds$ is very large in this region of kinetic energy, 
so we need to work with a
low average density material in order to have a reasonable electric
field strength, $|\vec{E}|$. 
Efficiency considerations lead to the use of a gas rather than a system
with alternating high and low densities, such as foils.  We could
tolerate larger electric fields in a system of foils.  However, the foils
will give larger energy loss fluctuations, resulting in problems with $\mu^-$
capture and muonium (bound state of $\mu^+$ and electron) formation.
In addition, a 
large fraction of the muons would be stopped in the foils and subsequently
lost.  We therefore pursue the use of a gas volume with a strong electric
field to extract the muons.

\item The lifetime of the $\mu^{\pm}$ can be written as

$$
v\tau \approx 0.1 \sqrt{T(eV)} \;\; {\rm m} \;\;\;\;\; T\ll m_{\mu}c^2 \; .$$

The muons should therefore only travel tens of centimeters at the low kinetic
energies to have a significant survival probability. They must then be
reaccelerated quickly to avoid loss due to decay.

\item A strong solenoidal field, $\vec{B}$, will be needed to guide the muons
until the beams are cooled. We cannot have  $\vec{E}$, in the cooling
cell, parallel to  
$\vec{B}$ or the muons will never get 
below the peak of the $dT/ds$ curve.  The muons 
will have typical kinetic energies well above the $dT/ds$ peak
when they
enter the gas volume. The electric field strength required to balance
$dT/ds$ 
at the low energies
would produce a strong acceleration of the muons at these high 
initial kinetic energies, such that the muons never slow down.  
We therefore consider an electric field transverse to the magnetic
field. In the cooling 
cell, the force on a muon is given by
$$\vec{F}=q(\vec{E}+\vec{v}\times\vec{B})-\frac{dT}{ds}\hat{v},$$
where $\hat{v}$ is a unit vector in the direction $\vec{v}$.
At the higher energies,
the muons follow the magnetic field lines with slow drift and pick up
minimal energy from the electric field.  Once the muons have slowed down,
however, the electric
force is no longer small compared to the magnetic force and the muons will
drift out of the volume at a definite Lorentz angle. In this way, the
cooling cell can be long in the initial direction of the beam and
short in the transverse direction in which the beam is extracted.

\item Large electric fields generally will result in breakdown in gases.
Having $\vec{E}\perp 
\vec{B}$ is expected to avoid this problem by limiting the maximum
kinetic energy an ionized electron can acquire and thus prevent charge 
multiplication.

\item Muonium (Mu) formation ($\mu^+ + {\rm Atom} \rightarrow {\rm Mu}
+ {\rm Atom}^+$)
is significant at  $\mu^+$ energies near the ionization peak.  In fact,
the muonium formation cross section dominates over the electron stripping
cross section in all gases except Helium~\cite{ref:Nakai}.  
This leads us therefore to choose Helium as a stopping medium (at least for
$\mu^+$).  

\item For $\mu^-$, a possibly fatal problem is the loss of muons 
resulting from muon capture, 
$\mu^- + {\rm Atom} \rightarrow \mu {\rm Atom} + e^-$. 
The cross section for this process has been calculated up to
kinetic energies of about $80$~eV~\cite{ref:hecohen,ref:h2cohen}, at which
point the 
cross section, for H$_2$ and Helium, is 
of order $10^{-17}$~cm$^2$, and falling rapidly.
Cross sections of order  $10^{-21}$~cm$^2$ are necessary for the cooling
described here to have significant efficiency for $\mu^-$. Large capture
cross sections extend to higher $T$ for higher atomic number $Z$.
To maximize the yield of $\mu^-$, we therefore
need to use a low $Z$ material and keep the kinetic energy as high as possible.
Helium or Hydrogen appear to be the best choices for the $\mu^-$ slowing
medium.

\item At the low energies used in Frictional Cooling, the
question of extracting muons from any gas cell becomes a significant
issue. The use of very thin windows as a possible solution will be
discussed in 
section~\ref{sec:windows}. 

\item Intense ionization from the slowing muons will produce a large
number of free charges which will screen the applied $\vec{E}$ field.
The speed at which this happens and the extent of the field reduction
have not been studied and will require further thought. It is briefly
discussed 
in 
section~\ref{sec:plasma}.

\end{itemize}

\subsection{Outline of a Scheme based on Frictional Cooling}

\begin{figure}
\begin{center}
\epsfig{file=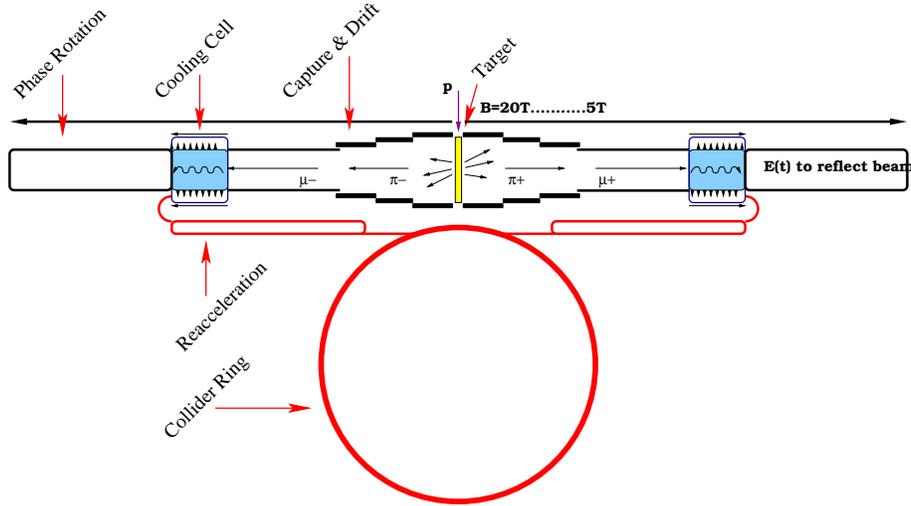,width=12.cm}
\end{center}
\caption[]{ Overview of a Muon Collider based on Frictional Cooling.
The different sections are not to scale.}
\label{fig:scheme}
\end{figure}

We have  investigated the scheme illustrated
in Fig.~\ref{fig:scheme}.
Muons of both signs are produced by
scattering an intense proton beam on a target located in a region with
a very strong 
solenoidal magnetic field ($20$~T, as in the Neutrino Factory study).
A drift region with a more moderate magnetic field, of the order of $5$~T,
allows the bulk of the pions to decay to muons.  
 The muons are then input into the cooling channel, which
consists of a cooling cell roughly $\sim 11~$m long. 
The cooling cell contains Helium gas (possibly H$_2$ for $\mu^-$), 
and an electric
field perpendicular to the magnetic field.  The electric field direction
is reversed periodically, as a function of the position along the axis
of the gas cell, to cancel 
the beam drift.  The muons stopped 
in the cell drift out at a characteristic angle dependent on
${\vec B}$ and ${\vec E}$. They are then extracted through thin
windows and reaccelerated. 

At the end of the drift
region, there is a correlation between the longitudinal momentum of the
muons and their arrival time.  This allows for a phase rotation, where
time varying electric fields are used to increase the number of muons at
lower momenta. In  contrast to schemes based on Ionization
Cooling, the phase rotation section follows the 
cooling cell, such that those muons already in the low energy regime,
which can achieve 
equilibrium in the medium, are not lost.  In the following, we
describe each of these elements and their simulations in more detail.

\section{Simulation of the Collider Front End}

The role of the target system for a Muon Collider is analogous to that
of a Neutrino Factory. One needs to generate a maximal number of pions
with an intense proton beam and then capture and guide them into a
channel where they can decay to muons. The muons from the pion decays
can then  be cooled, 
reaccelerated  and stored 
in a ring for subsequent
injection into a collider. However, the evaluation criteria for optimizing the
pion yield is different for the Neutrino Factory from that of a Muon
Collider based on Frictional Cooling. By virtue of the low energy
required for Frictional Cooling, the target system is
optimized for the yield of low energy pions.
The starting point was taken from the Feasibility Study II
for the Neutrino Factory~\cite{ref:STUDYII}. We then
modified the design to fit our needs.

The following target yield studies were performed using the
MARS Monte Carlo code~\cite{ref:MARS}. MARS performs fast, inclusive
simulations of three-dimensional hadronic and electromagnetic
cascades, and performs muon and low energy neutron transport in
material, in the 
energy range of a fraction of an eV up to 30~TeV. The subsequent
transport of produced particles in the target area and decay channel
was simulated using the 
GEANT 3.21 package~\cite{ref:GEANT}. The cooling and reacceleration
simulation was performed with custom written code.

\subsection{Targetry}

The initial target geometry studied consisted of the Study
II~\cite{ref:STUDYII} target, 
in which the proton beam impinged a target under a small angle.
A strong solenoidal field is used to capture pions in a wide phase space.
The extraction of the
pion beam takes place in the longitudinal direction (the direction of
the incident proton beam) in the Neutrino
Factory scheme. 

\begin{figure}
\begin{center}
\epsfig{file=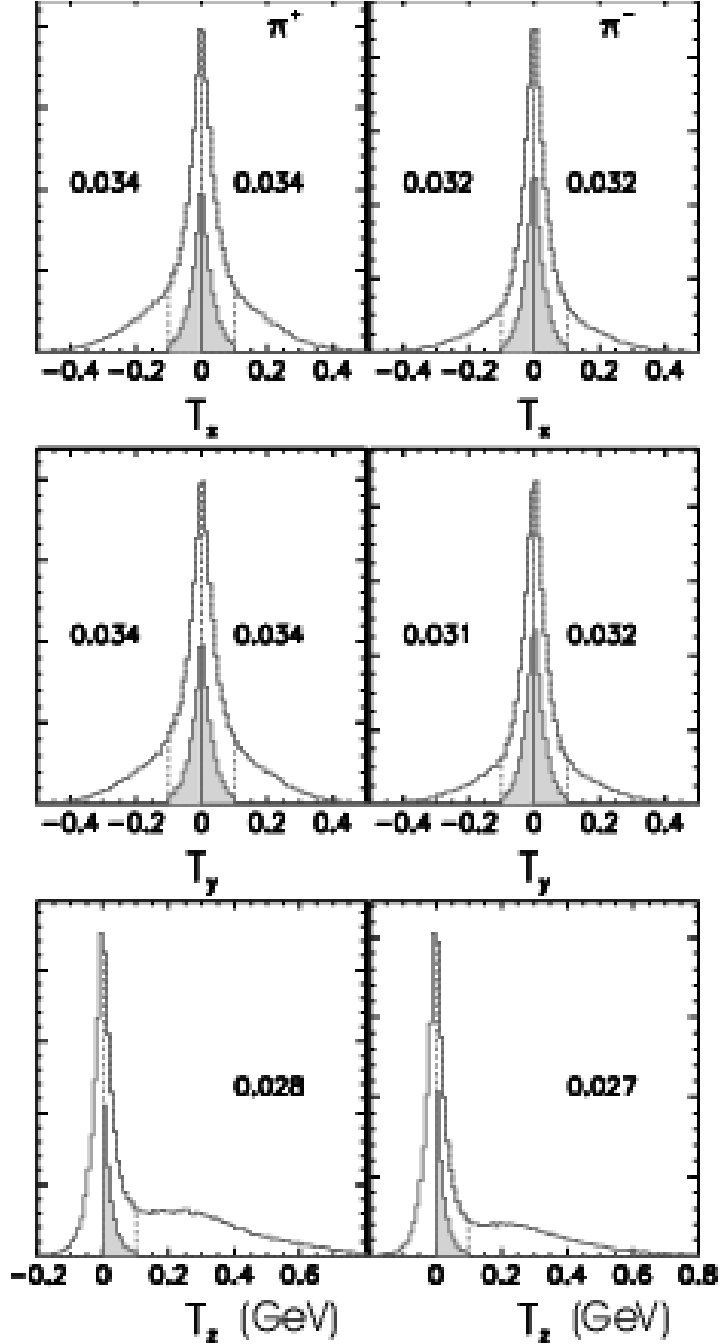,height=18cm}
\caption{\label{fig:signedT} Signed kinetic energy (in units of GeV) of
$\pi^+$ (left) and $\pi^-$ (right)  produced at 
the target, for a 2~GeV proton beam on a $0.75~{\rm cm}\times 30~{\rm
cm}$ Copper target. $T_{x,y,z}$ is defined as $\frac{P_{x,y,z}}{|P_{x,y,z}|}T_{x,y,z}$. The shaded histograms correspond to
those pions 
with a kinetic energy less than 100~MeV. The vertical scale is
arbitrary. The numbers on the plots indicate the number of pions per 1~GeV
proton on target within the shaded regions. }
\end{center}
\end{figure}

This is not ideal for Frictional Cooling because the
highest momentum muons, which result from decays of 
the
highest momentum pions, would not be cooled. Frictional Cooling
requires the optimization of the pion yield at low energies.
Figure~\ref{fig:signedT} shows 
the signed kinetic energy of the produced $\pi^\pm$ from the target.
The z-direction is the 
longitudinal or proton beam  direction.
The low energy
pion yield transverse to the target is larger than that in the
longitudinal direction. Moreover, there are relatively equal yields
for $\pi^+$ and 
$\pi^-$. This allows the possibility to develop a symmetric $\mu^+$
and $\mu^-$ 
machine and hence cool both signs at the same time. We have therefore
chosen to extract the pions produced transverse to the target. The
magnetic field is therefore perpendicular to the direction of the
proton beam.

\subsubsection{Layout of Target Area}

The pion capture mechanism
consists of a solenoidal magnetic field channel 
starting with B=20 T near the target and falling adiabatically to 5 T at
18~m.
Particle capture/extraction takes place transverse to the target or
direction defined by the proton beam. 

\begin{figure}
\begin{center}
\epsfig{file=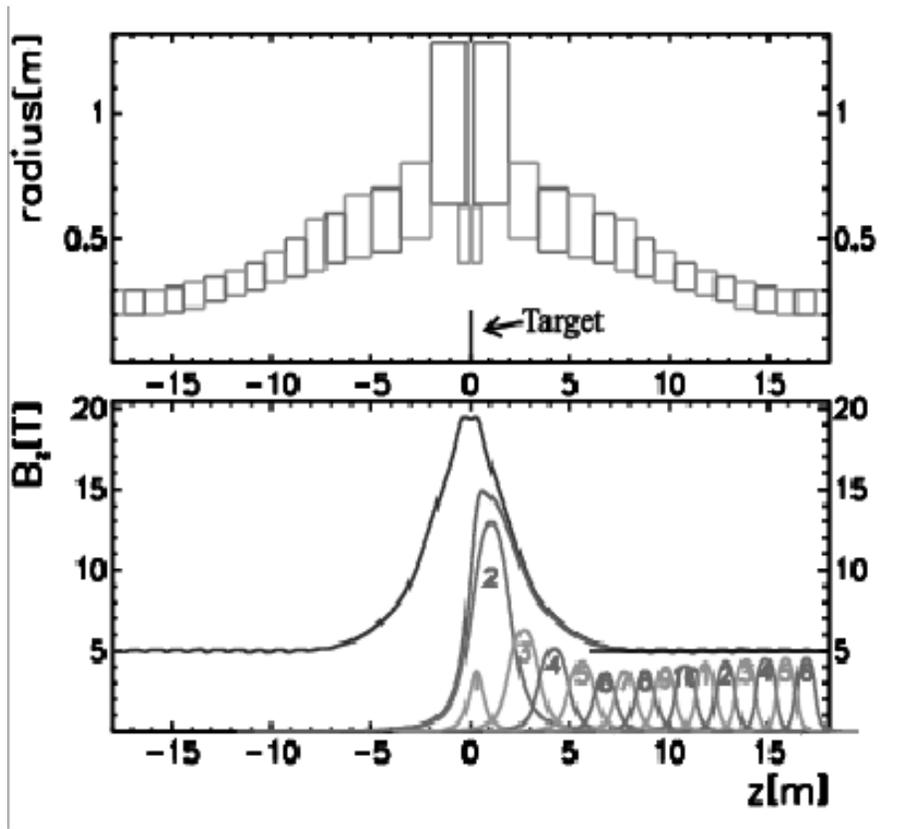,width=12cm}
\caption{\label{fig:magfield}
The upper plot shows schematically the arrangement, in cross section, of the
solenoids used to create the tapered solenoidal field. The lower plot
shows the contributions from individual solenoids and their
cumulative field along the z axis. The target is represented by the
solid line at z=0.}
\end{center}
\end{figure}

The magnetic field is symmetric with a gap for the
non-interacting proton beam, and forward produced particles, to pass
through. 
Figure~\ref{fig:magfield} shows the dimensions of the solenoid coils and
magnetic field profile. 
Note that the z-direction is redefined in this
figure to
the direction of extraction which is transverse to the target.
The proton beam trajectory in this strong
field has not been simulated in 
detail.

\subsubsection{Optimization of Yields}
\label{sec:yields}

\begin{figure}
\begin{center}
\epsfig{file=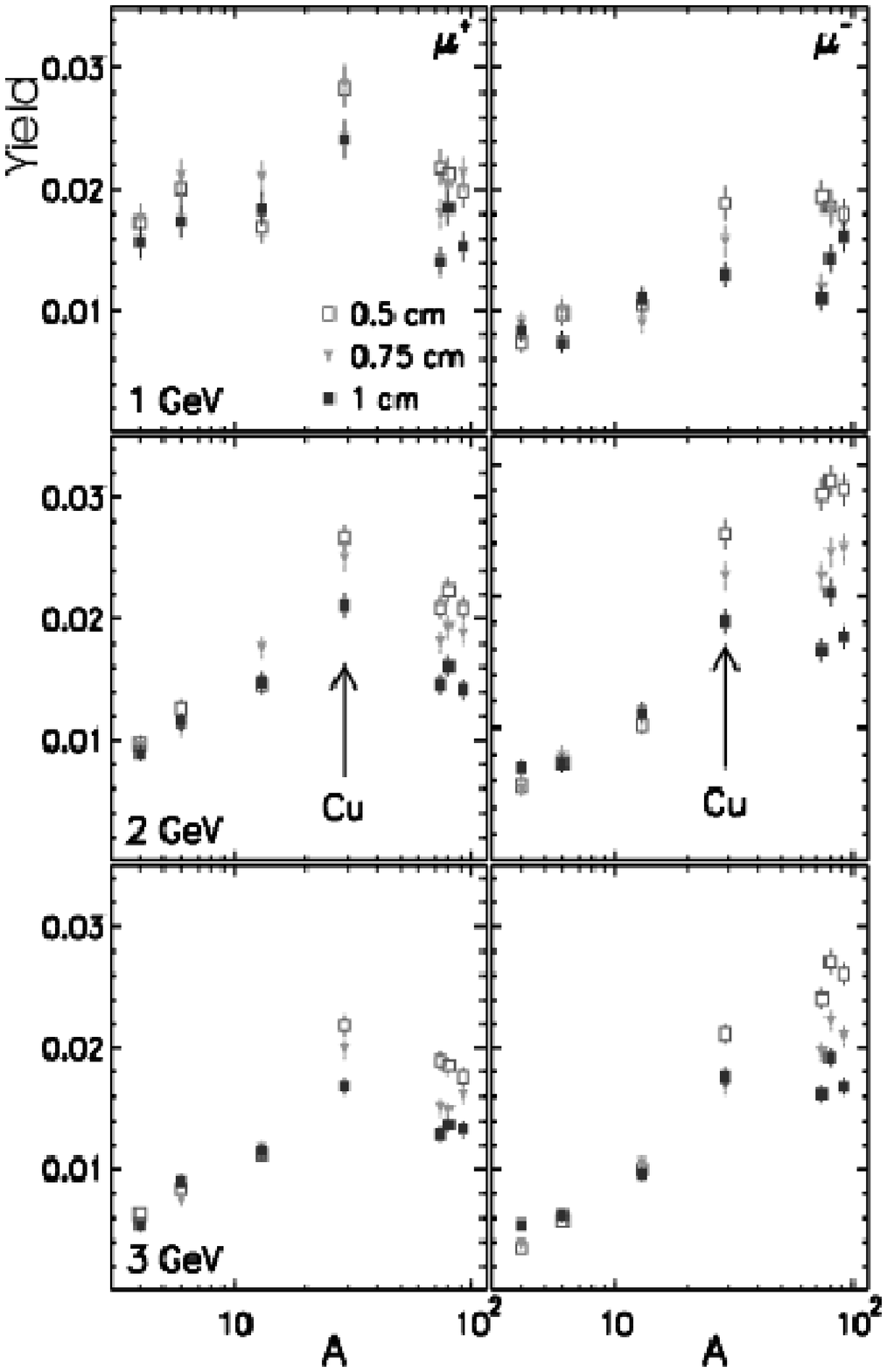,width=12cm}
\caption{\label{fig:yieldA}Yields  of $\mu^+$ (left) and $\mu^-$
(right), in units of 
number of $\mu$ per 1~GeV p, as a function of the 
atomic mass of the target material for different target thicknesses as
denoted by the symbols. The rows correspond to different proton
driver beam energies; 1~GeV, 2~GeV and 3~GeV as denoted in the figures.}
\end{center}
\end{figure}

An optimization of the composition and geometry of the target was
performed to find the maximal pion 
yield, which would produce {\it coolable} muons. Coolable muons were
defined as those with a kinetic energy less than 120~MeV. The target
material, geometry and the energy of the proton
driver, were varied to find the optimum yield. The results of this
study are summarized 
in Fig.~\ref{fig:yieldA}.

The choice of a Copper target resulted
in the optimal 
yield. A Tungsten target produced similar performance and may be more
desirable in terms of stability and target lifetime.

\begin{figure}
\begin{center}
\epsfig{file=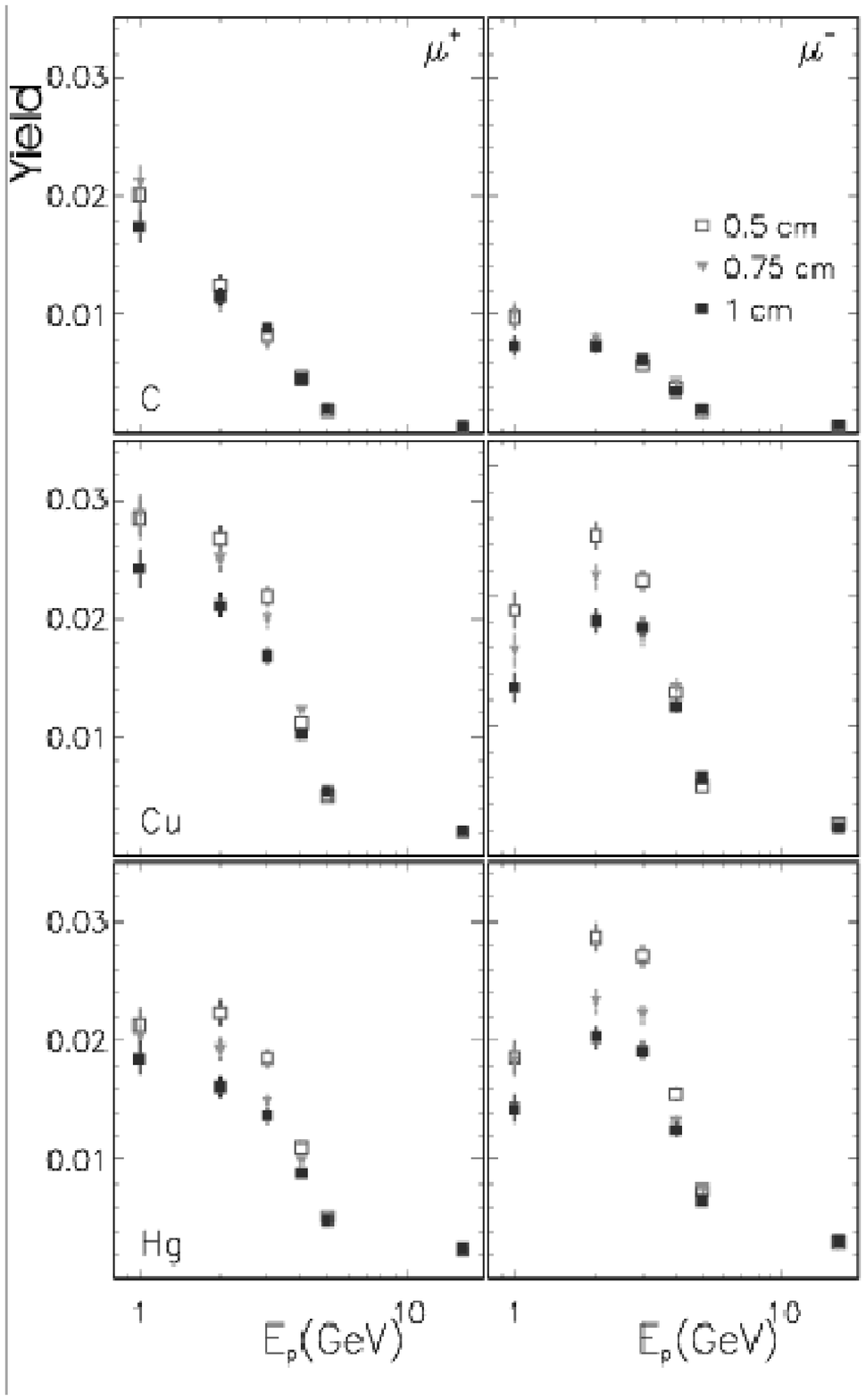,width=12cm}
\caption{\label{fig:yieldep}Yields  of $\mu^+$ (left) and $\mu^-$
(right), in units of number of  
$\mu$ per 1~GeV p, as a function of
the proton driver beam energy for different target thicknesses as
denoted by the symbols. The rows correspond to different target
materials; Carbon, Copper and Mercury as denoted in the figures.}
\end{center}
\end{figure}

Figure~\ref{fig:yieldep} shows the effect of the proton driver beam energy
on the yield for various target species and thicknesses. 
A proton beam
energy of 2 GeV results in the highest yield and roughly equal yield 
for both signs of pions. This is important in the development of a
$\mu^+\mu^-$ collider.

\subsection{Drift Region}

\begin{figure}
\begin{center}
\epsfig{file=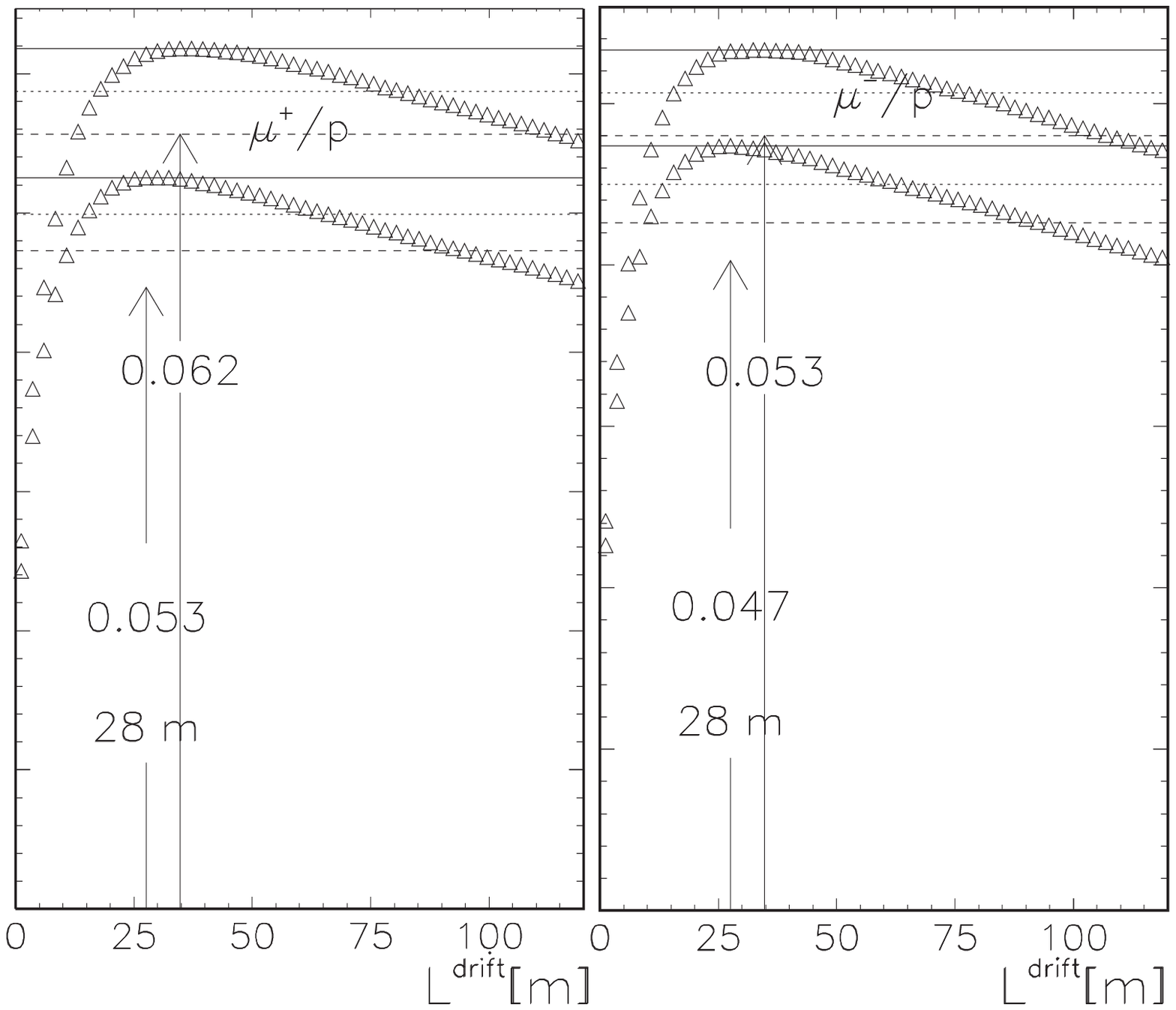,width=12cm}
\caption{\label{fig:drift} Yield as a function of decay channel length
$L^{drift}$ for $\mu^+$ (left)  and $\mu^-$ (right), in units of
number of $\mu$ per 1~GeV p. The lower symbols indicate the number of
$\mu$ per 1~GeV p with $T_z<120~$MeV.}
\end{center}
\end{figure}

The solenoidal field is tapered down from 20~T at the
target to 5~T in order to reduce the amount of stored energy of such
a section and to reduce cost. We have used a similar design to that
given in \cite{ref:STUDYII}. The length over which the pion beams
are allowed to drift and decay was then optimized (for the muon
yield). Figure~\ref{fig:drift} 
shows the results of this study.
The optimum drift length 
is taken to be the distance which gives the maximum $\mu^+/p$ yield
for $T_{\mu^+}<120$~MeV and is found to be
28 m.

\subsection{Summary of Target and Drift regions}

In summary, the optimization of the front end of a muon Frictional
Cooling scheme resulted in the following scenario:
\begin{itemize}
\item a 2~GeV proton driver
\item a Copper target 30~cm in length and 0.5~cm thick
\item a drift length of 28~m
\end{itemize}
The yield at the end of the drift
of such a front end is $0.062~\mu^+$ per 2 GeV proton and
$0.054~\mu^-$ per 2 GeV proton.

\subsection{Phase Rotation Section}

\begin{figure}
\begin{center}
\epsfig{file=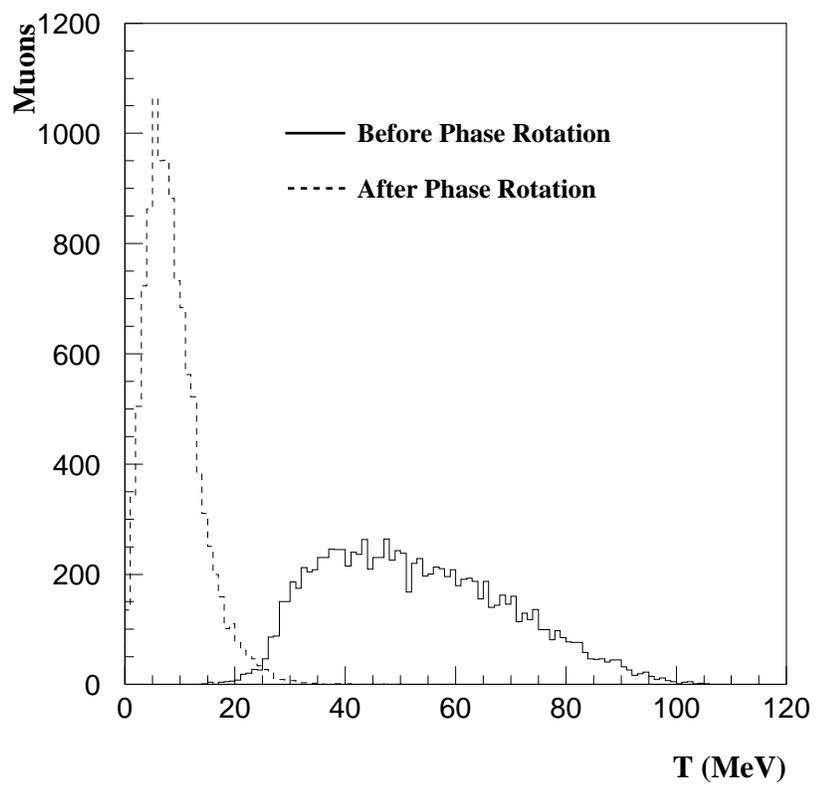,width=12.cm}
\end{center}
\caption{The kinetic energy, $T$, distributions for a $\mu^+$ beam before
and after the phase rotation section. Both before and after
distributions correspond to muons which are eventually cooled. }
\label{fig:phaserotate}
\end{figure}

At the end of the cooling cell, there is a correlation between the
longitudinal momentum of the muons which survive and the arrival time.
This allows 
for a phase rotation to take place, where time varying electric
fields are used to increase the number of muons at lower momenta. A
simple ansatz was used for the phase rotation section and consisted
of a flat 5~MV/m field for some time $t_1$, after which the field went
linearly to zero at a time $t_2$. These parameters were optimized
($t_1=100$~ns and $t_2=439$~ns) to
obtain the maximum number of muons/proton coming from the 
front end. The effect of the phase rotation on the kinetic energy
distribution of the muons is shown in Fig.~\ref{fig:phaserotate}.

\section{Cooling Cell}

The simulated  cooling cell is a cylinder filled with either Helium
(for $\mu^+$ 
or $\mu^-$) or Hydrogen (for $\mu^-$) gas. The axis 
of the cylinder defines the z-direction (or longitudinal
direction) and corresponds to the direction of the muon beam. The
entrance windows at 
either end must pass $T\leq 50$~MeV muons with high efficiency. The
extracted muons come out transversely and have low energy. The windows
for the transverse extraction must have high efficiency for $T\leq
1$~keV muons.

The
gas cell consists of one  continuous volume, 11~m in length. The cell
is located immediately after the drift region to capture those muons
which are already slow enough to be stopped in 11~m of gas. The
bulk of the muons at higher energies pass unaffected through the gas 
cell into the phase rotation section. The phase rotation
reflects the beam using time changing electric fields
and is optimized to produce a maximum number of muons at low kinetic
energies such that they can be stopped in the gas cell. 

The density of Helium used in the simulation for $\mu^+$ was
$1.25 \cdot 10^{-4}$~g/cm$^3$, which corresponds to roughly 0.7~atm at
STP. For $\mu^-$, 0.3~atm of He or H$_2$ was used,
which correspond to densities of $5.35\cdot 10^{-5}$ and $2.5 \cdot
10^{-5}$~g/cm$^3$ respectively.

The window thicknesses for the extraction of low energy muons varied
from 0-20~nm of Carbon. These windows are conceived as grids upon
which a Carbon film is deposited. The grid would have a large open
area ($>80\%$), and this loss in efficiency has not been simulated.

\begin{figure}
\begin{center}
\epsfig{file=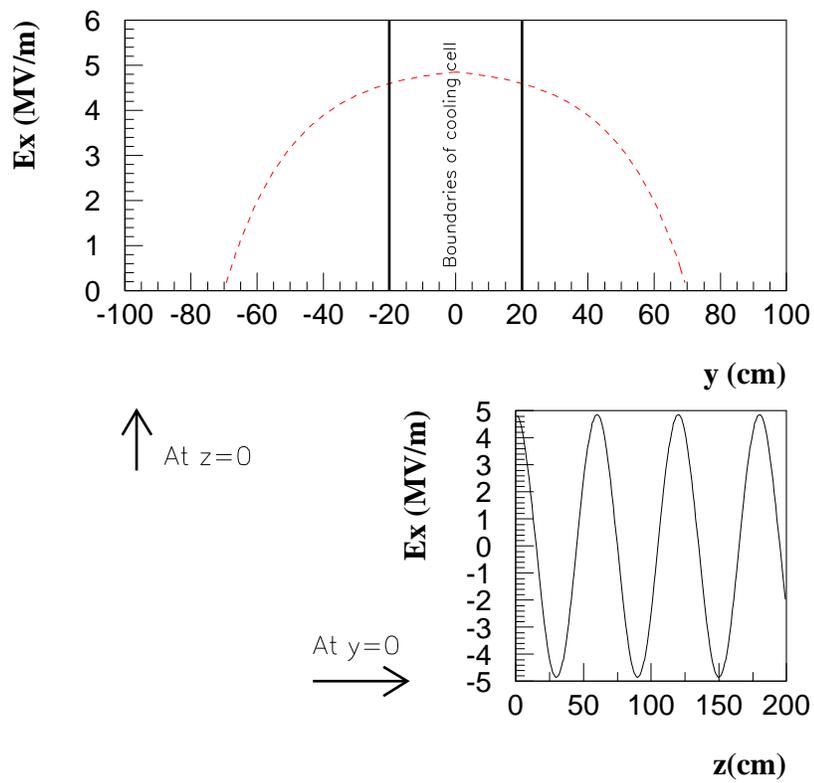,width=12cm}
\caption{Variation of E$_x$ inside the cooling cell region. The upper
plot indicates the variation of E$_x$ as a function of y at z=0 and
the lower plot shows the variation of E$_x$ as a function of z at y=0.}
\label{fig:exfield}
\end{center}
\end{figure}
The entire cooling and phase rotation sections were contained inside a
uniform 5~T solenoidal field and the crossed electric field strength
was taken to be 5~MV/m inside the gas cell section. The electric
field oscillated as a cosine function every 30~cm inside the gas cell
and was exponentially damped to zero at a radius 70~cm. The E$_x$ field
profile is defined as follows,
$$E_x=E_0 \cdot cos(z [{\rm cm}]\pi/30.)\cdot(1-exp(-(|y[{\rm cm}]|-70)/20)),$$
and can be seen in Fig.~\ref{fig:exfield}. 
The technical
feasibility of such a field configuration has not yet been
investigated.

\subsection{Tracking}

The physics processes are described in more detail in the next section - 
we focus here on the technical aspects of the tracking.   
Muon decay and muon capture were not simulated but were taken
into account by calculating the probability 
that the muons survived these processes.  

The tracking of muons proceeded as follows:
\begin{itemize}
\item
The position of the muon was used to determine the local value of
the electric and magnetic fields, as well as the local material and density.
\item
Given the muon momentum and the medium in which it was located, the mean free
path to the next nuclear scatter was calculated.  
The mean free path was then used to 
generate a track step, $D$.
\item
$D$ was then compared to the maximum step 
size, $d$, (typically 10~$\mu$m), set to preserve accuracy in tracking.  
If $D> d$ then $D$ was broken up into 
several steps of size $\le d$.
\item
The Runge-Kutta-Merson approach was then used to update the position and
momentum of the muon according
to the Lorentz force equation for each step using 
the CERN routine~\cite{ref:CERNLib} DDEQMR.  The force law included
the continuous energy loss from scattering on electrons via:
$$\vec{F} = \frac{d \vec{P}}{d t} = q\vec{E} + q \vec{v}\times\vec{B} - 
 \frac{d T}{d s}\hat{v},$$
where $T$ is the muon kinetic energy and $s$ is the path length.  The total
time was updated at each step, and the time interval in the step was
used to update the decay survival 
probability.  For negative muons, the capture survival probability was
also updated based on the step length and kinetic energy.
\item
Once the full distance $D$ was reached, a 
scattering angle was generated according to the relevant differential
cross section.  The momentum components of the muon were then updated
appropriately.
\end{itemize}
This procedure was followed until either the muon left the cooling cell
volume or one of the survival probabilities fell below $0.001$.  The
tracking code was tested extensively and was found to give very
accurate results.  Examples of muon trajectories are given in 
section~\ref{sec:trajectories}.

\subsubsection{Simulation of Physics Processes}

\begin{figure}
\begin{center}
\epsfig{file=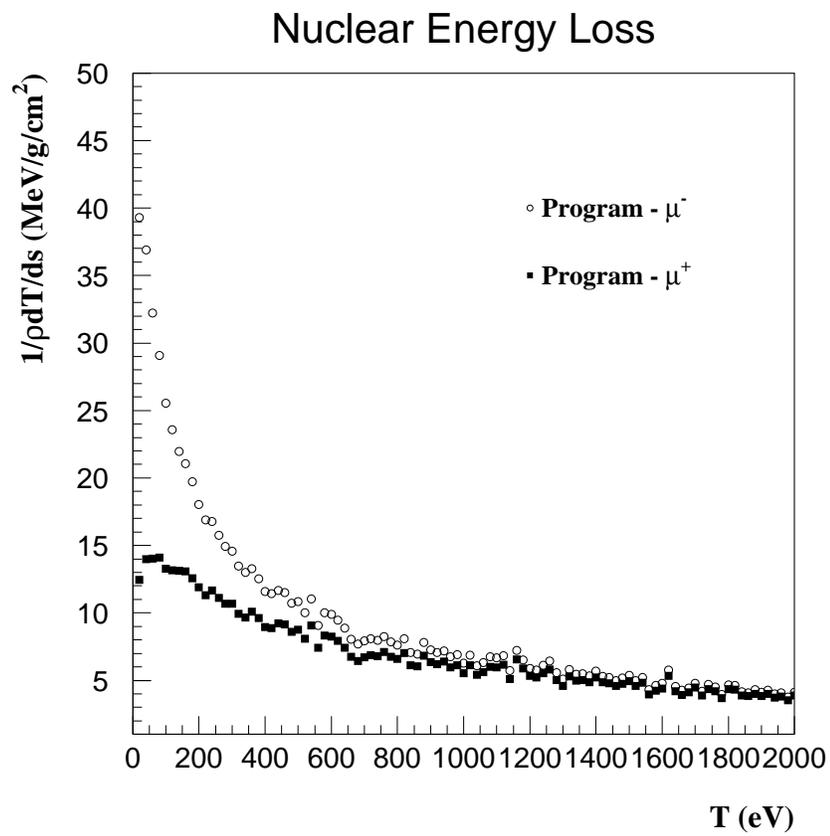,width=12.cm}
\caption{Energy loss from nuclear scattering or nuclear stopping power
for $\mu^+$ and $\mu^-$ produced using
the stand alone Monte Carlo.}
\label{fig:compmus}
\end{center}
\end{figure}

\begin{figure}
\begin{center}
\epsfig{file=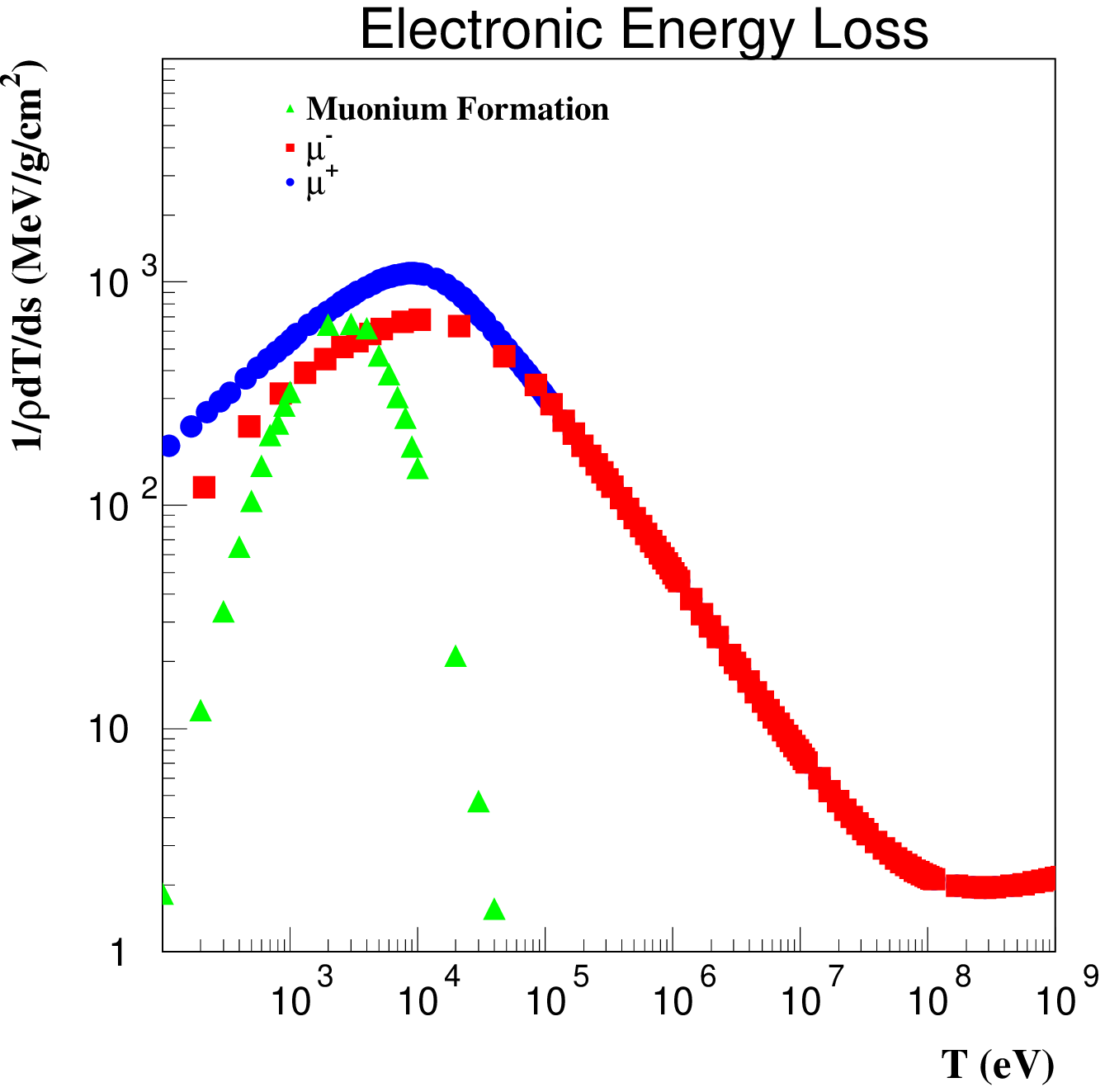,width=12cm}
\caption{The electronic energy loss for $\mu^+$ and $\mu^-$. For
comparison, the simulated energy loss due to muonium formation (see
text) is also shown.} 
\label{fig:barkas}
\end{center}
\end{figure}

Frictional Cooling cools muon beams to the limit of nuclear
scattering. A detailed simulation was therefore performed where all
large angle nuclear scatters were simulated. The differential
distributions and mean free paths for $\mu$-nucleus scattering were
calculated in two different ways depending on the energy regime. A
quantum mechanical (Born Approximation) calculation was used for the
scattering cross 
section at high kinetic energies ($T_{\mu}>2$~keV) and a classical
calculation was used at lower kinetic energies. The screened Coulomb
potential used in the Born and classical calculations had the form
$$V=\frac{4e^2}{r}~exp(\frac{-r}{a}),$$
where $a$ is the screening length.
For the classical
calculation the procedure of Everhart et
al.~\cite{ref:Everhart} was followed. The differential cross section
was calculated by scanning in impact parameter and evaluating the
scattering angle at each impact parameter.  From the
differential cross section, a mean free path for scattering
angles greater than a cutoff (0.05~rad) was found, and
scatters were then generated according to the differential cross
section. This method reproduces the 
energy loss from nuclear scatters tabulated by NIST~\cite{ref:NIST}
for protons. The simulation results for
$\mu^+$ and $\mu^-$ are
shown in Fig.~\ref{fig:compmus}. 
For $\mu^+$, the electronic energy
loss is taken 
from the NIST tables. The suppression for $\mu^-$ (Barkas
effect~\cite{ref:Barkas}) was 
parameterized 
from the results in~\cite{ref:Agnello}.
The electronic energy loss was treated as a continuous process.
The electronic energy losses for
$\mu^+$ and $\mu^-$ are shown in Fig.~\ref{fig:barkas}. 
This difference is
significant below the ionization peak, which is expected to be due in
part to muonium formation.

\begin{figure}[htb]
\begin{center}
\begin{minipage}[t]{0.48\textwidth}
\epsfig{file=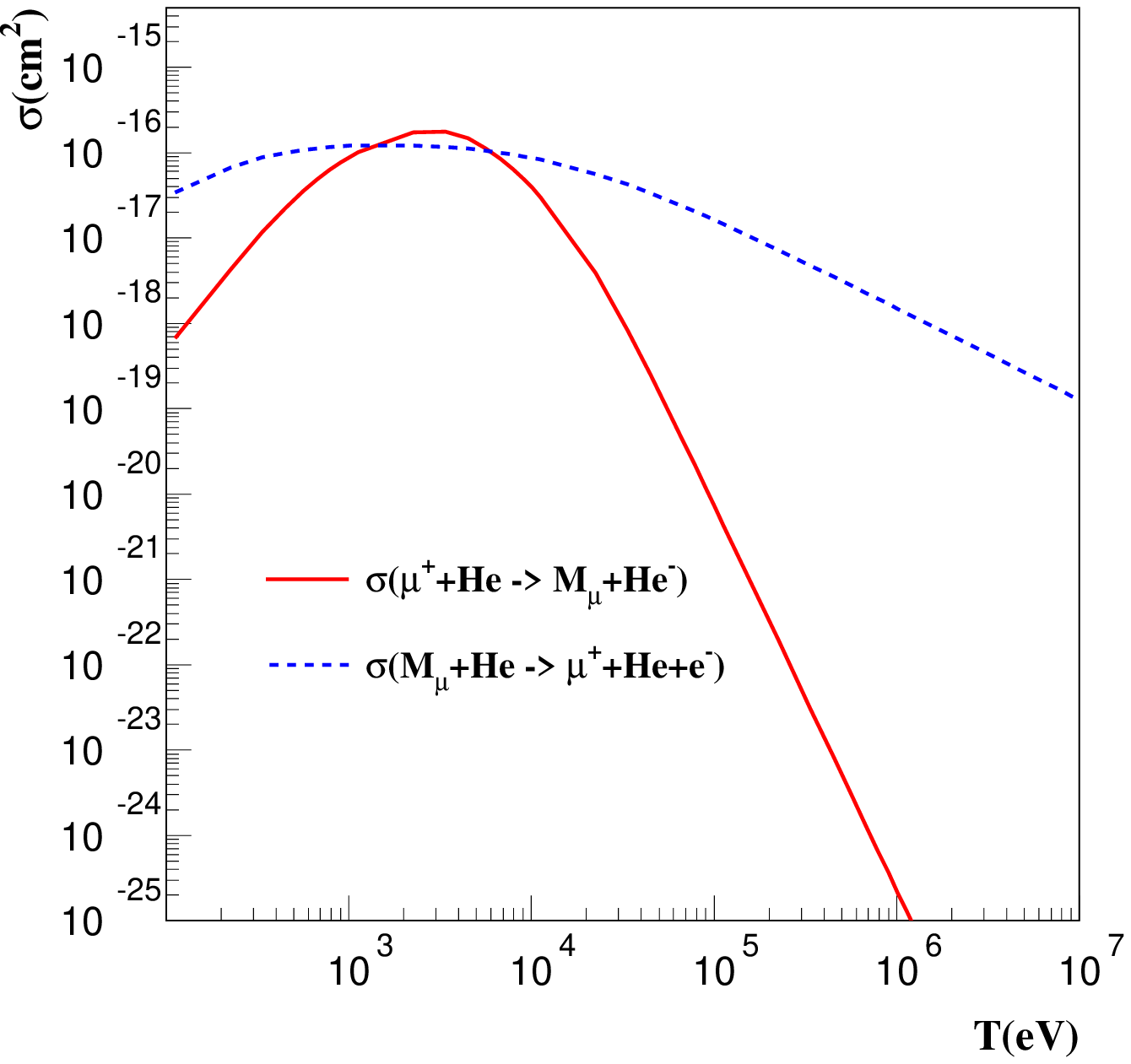,width=0.95\textwidth}
\end{minipage}
\begin{minipage}[t]{0.48\textwidth}
\epsfig{file=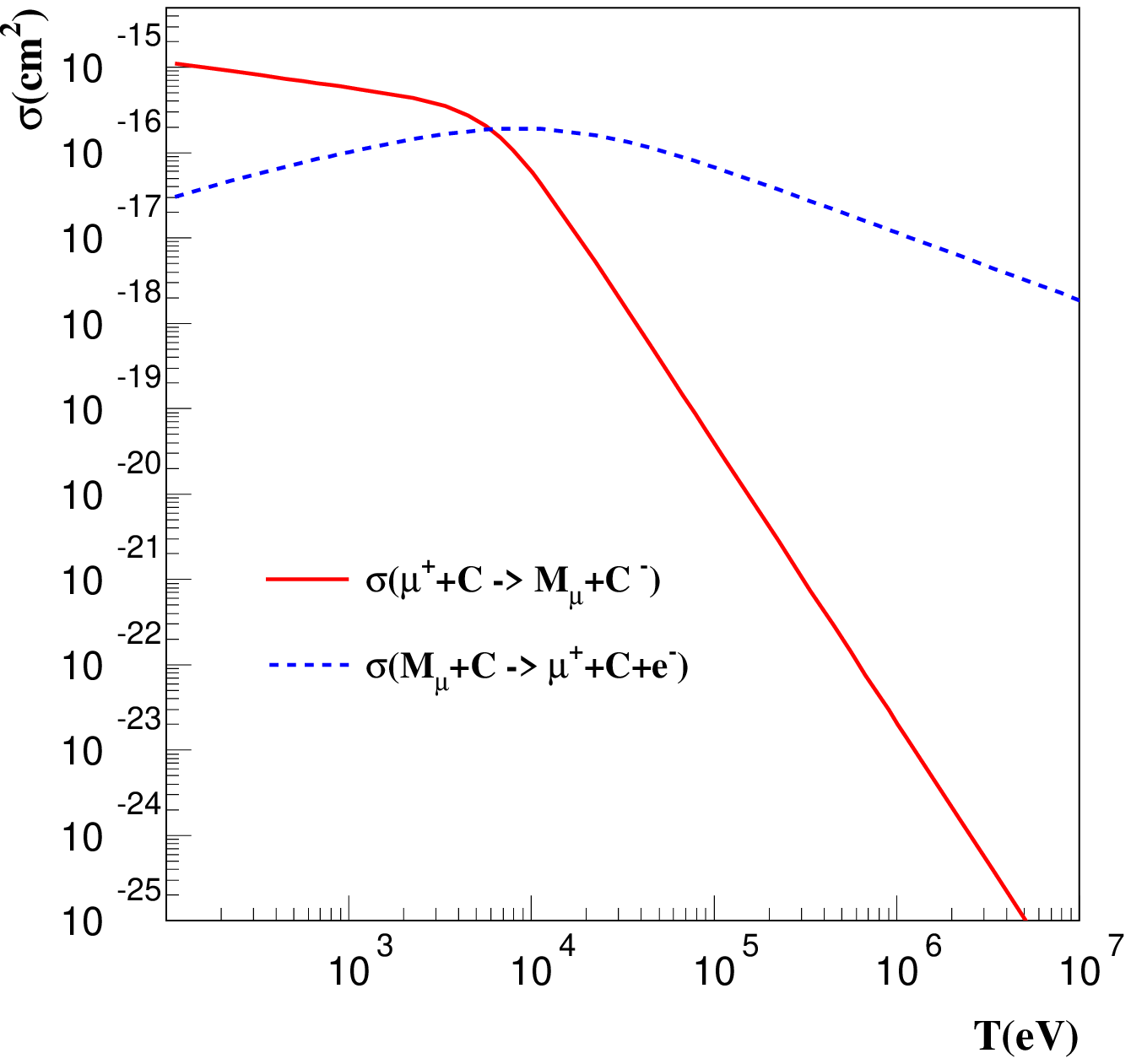,width=0.95\textwidth}
\end{minipage}
\caption{The muonium formation, $\sigma(\mu^+ +{\rm A}\rightarrow
M_\mu+{\rm A}^-)$, and muonium ionization, $\sigma(M_\mu +{\rm A}\rightarrow
\mu^+ + {\rm A}+ e^-)$, cross sections
for Helium (left) and Carbon (right)~\cite{ref:Nakai}.}
\label{fig:hec}
\end{center}
\end{figure}

To simulate the effect of muonium formation in the tracking, an
effective charge was used, as 
given by $\sigma_I/(\sigma_F+\sigma_I)$, where
$\sigma_I$ is the cross section for muonium ionization and
$\sigma_F$ is the cross section for muonium  formation (see
Fig.~\ref{fig:hec}).  

\begin{figure}
\begin{center}
\epsfig{file=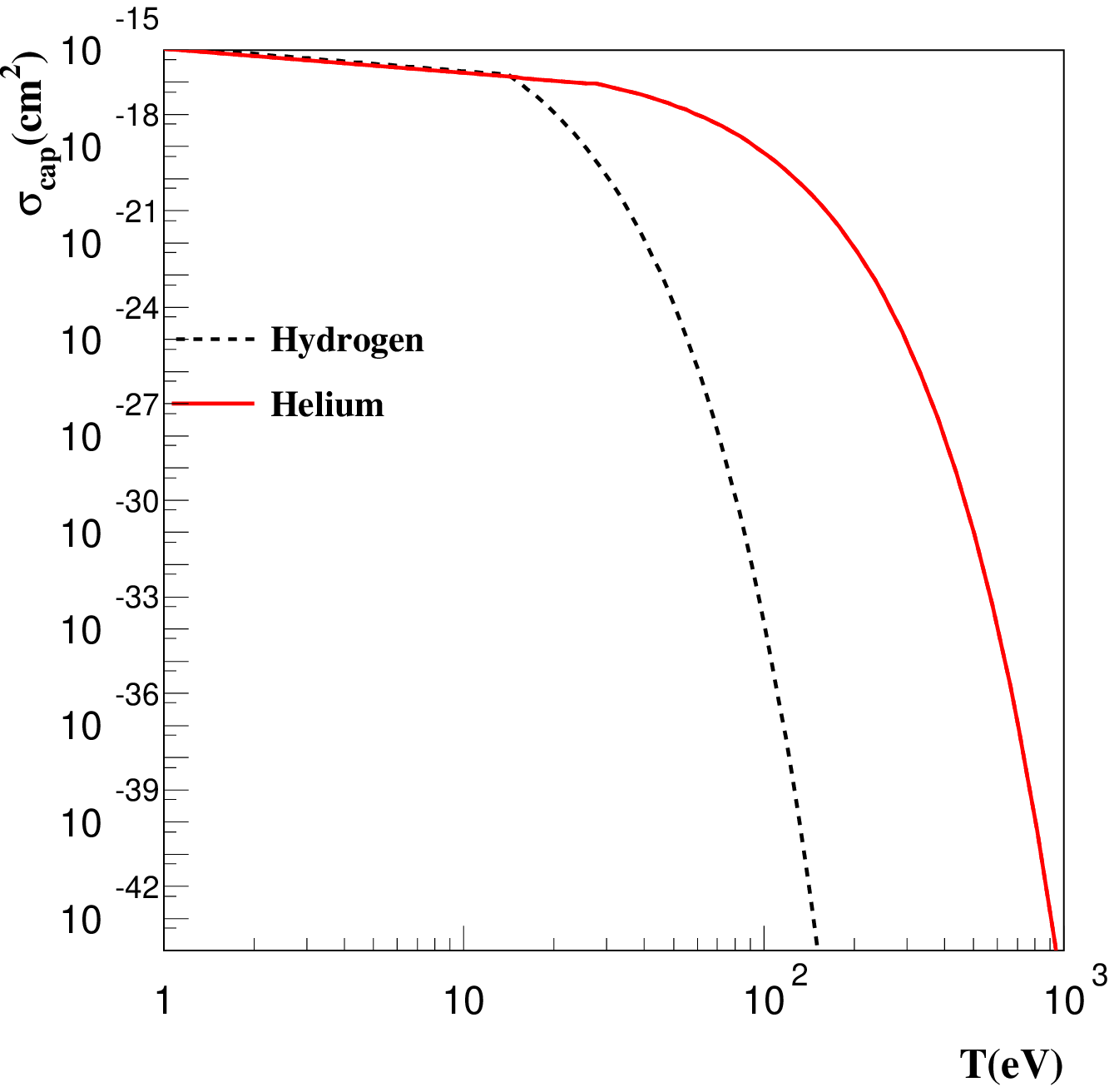,width=12cm}
\caption{Parameterization of $\mu^-$ capture cross section,
$\sigma_{\rm cap}$, in
Hydrogen and Helium
used in the simulation as a function of the kinetic
energy of the $\mu^-$, $T$. }
\label{fig:mucap}
\end{center}
\end{figure}

Negative muon capture was parameterized from calculations of
Cohen~\cite{ref:hecohen,ref:h2cohen}
and included
in our simulation. The calculations only extend up to 80~eV. Beyond this, a
simple exponential fall off with kinetic energy was assumed. The
parameterization is plotted in Fig.~\ref{fig:mucap}.

\subsection{Trajectories}
\label{sec:trajectories}

\begin{figure}
\begin{center}
\epsfig{file=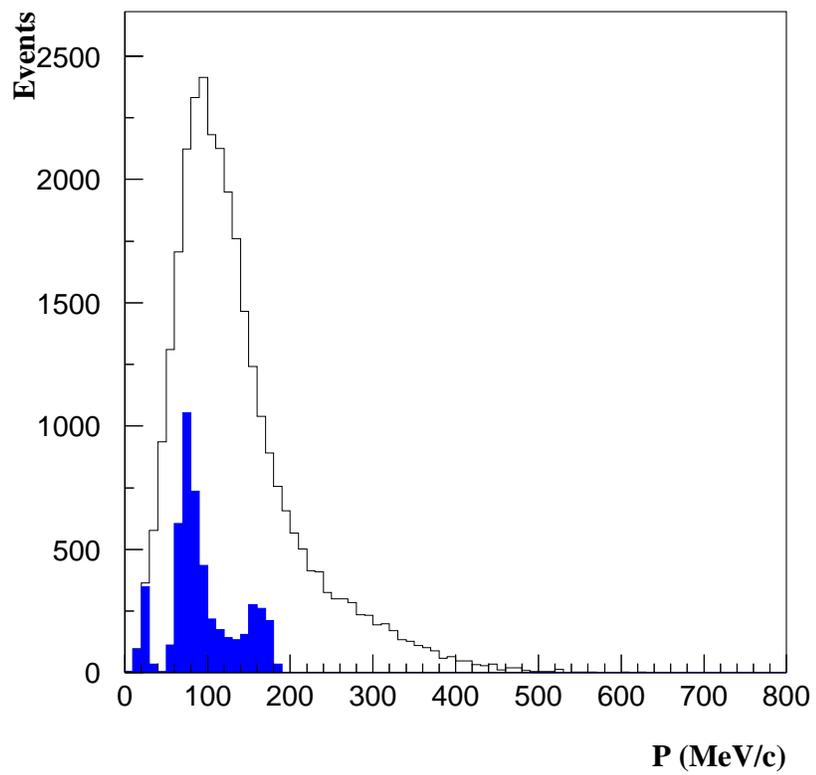,width=12cm}
\caption{The momentum, $\vec{P}$, distribution of muons entering the
cooling section (histogram). The  distribution of those which result
in cooled muons is shaded.}
\label{fig:distp}
\end{center}
\end{figure}

\begin{figure}
\begin{center}
\epsfig{file=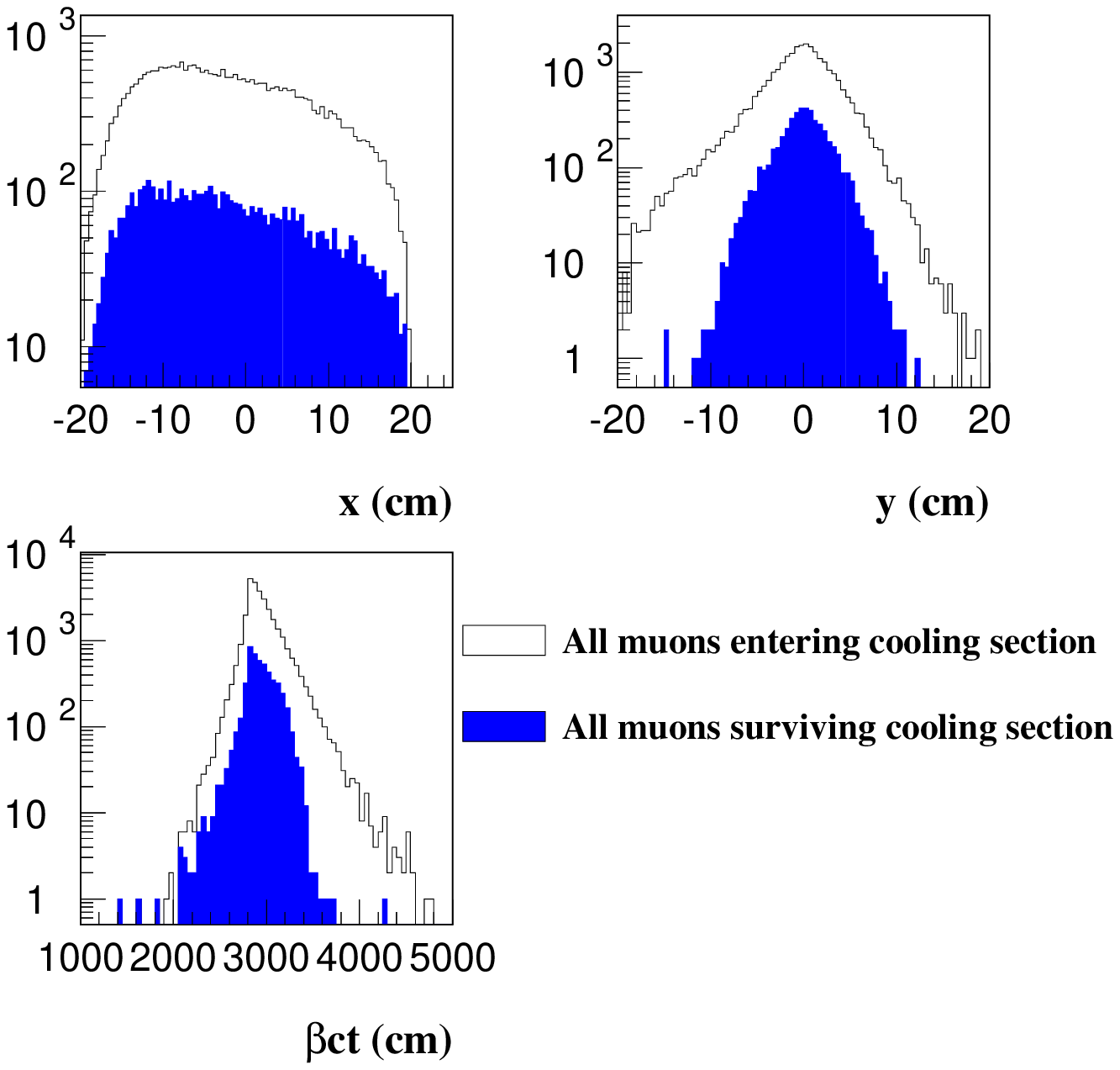,width=12cm}
\caption{The transverse and longitudinal spatial  distributions of muons
entering the cooling section (histogram). The distributions of those
which result in cooled muons is shaded. The top plots indicate the
transverse, x and y, spatial distributions and the bottom plot shows the
longitudinal spatial distribution expressed as $\beta c t$, where
$\beta c$  is the muon velocity and $t$ is absolute  time. }
\label{fig:distxyz}
\end{center}
\end{figure}

The momentum and space distributions of muons at the end of the decay channel
are displayed in Fig.~\ref{fig:distp} and Fig.~\ref{fig:distxyz}. The
full histograms show the input  
distributions, while the shaded histograms show the distributions of muons
which survive.  From these distributions,
one can see that this cooling channel does not cool muons with a
momentum greater than 200~MeV/c.  In the absence of electric and
magnetic fields the range, in meters, for a $\mu^+$ in He gas
with density  
$\rho=1 \cdot 10^{-4}$~gm/cm$^3$ was found to be
$$R~{\rm [m]}=1.2 \cdot 10^{-4} (|\vec{P}|~{\rm [MeV/c]})^{3.35},$$
with the momentum given in units of
(MeV/c).  The cooling cell has a length of 11~m. Due to the 
transverse momentum of the muon, the total path length in the cell is
on average 20~m. Therefore, muons with $\vec{P}\le 35$~MeV/c will be
stopped in the 
gas cell.  This can either happen on the first pass (first peak in
Fig.~\ref{fig:distp}), or after the phase 
rotation/reflection (second and third peaks in Fig.~\ref{fig:distp}).  For
the maximum momentum, muons require about $300$~ns 
to reach the equilibrium energy, $T_{eq}$. The time required to drift
transversely out of the cell depends on where the muon reached
$T_{eq}$.  Taking as nominal values $T_{eq}=500$~eV and  a drift
distance of 20~cm (cell radius) yields an additional time of $200$~ns.
Additional flight 
times appear for the muons which are reflected by the phase rotation section
and in the initial reacceleration (described below).  The  time 
required for cooling the muons and extracting them is therefore of order
$1$~$\mu$s.

\begin{figure}[htb]
\begin{center}
\begin{minipage}[t]{0.48\textwidth}
\epsfig{file=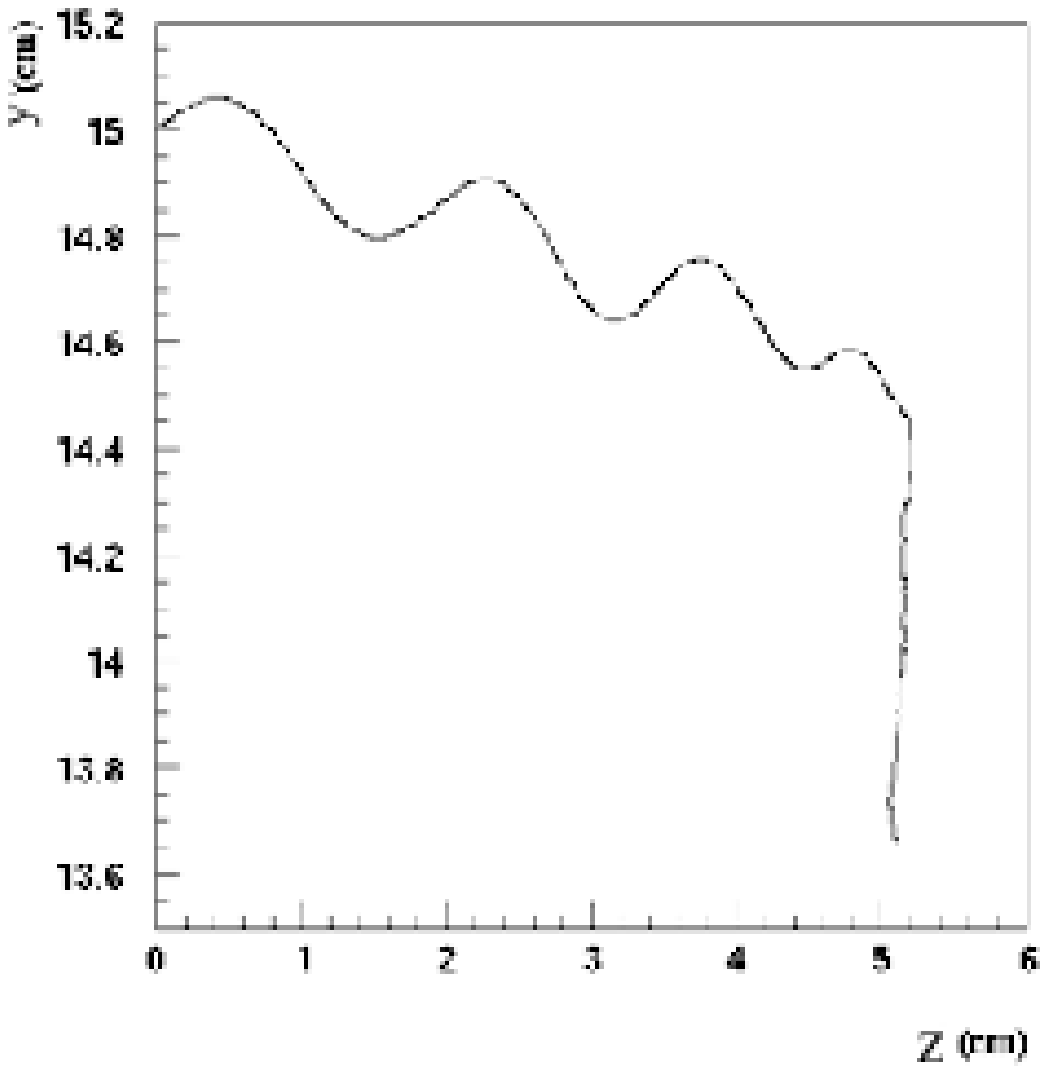,width=0.95\textwidth}
\end{minipage}
\begin{minipage}[t]{0.48\textwidth}
\epsfig{file=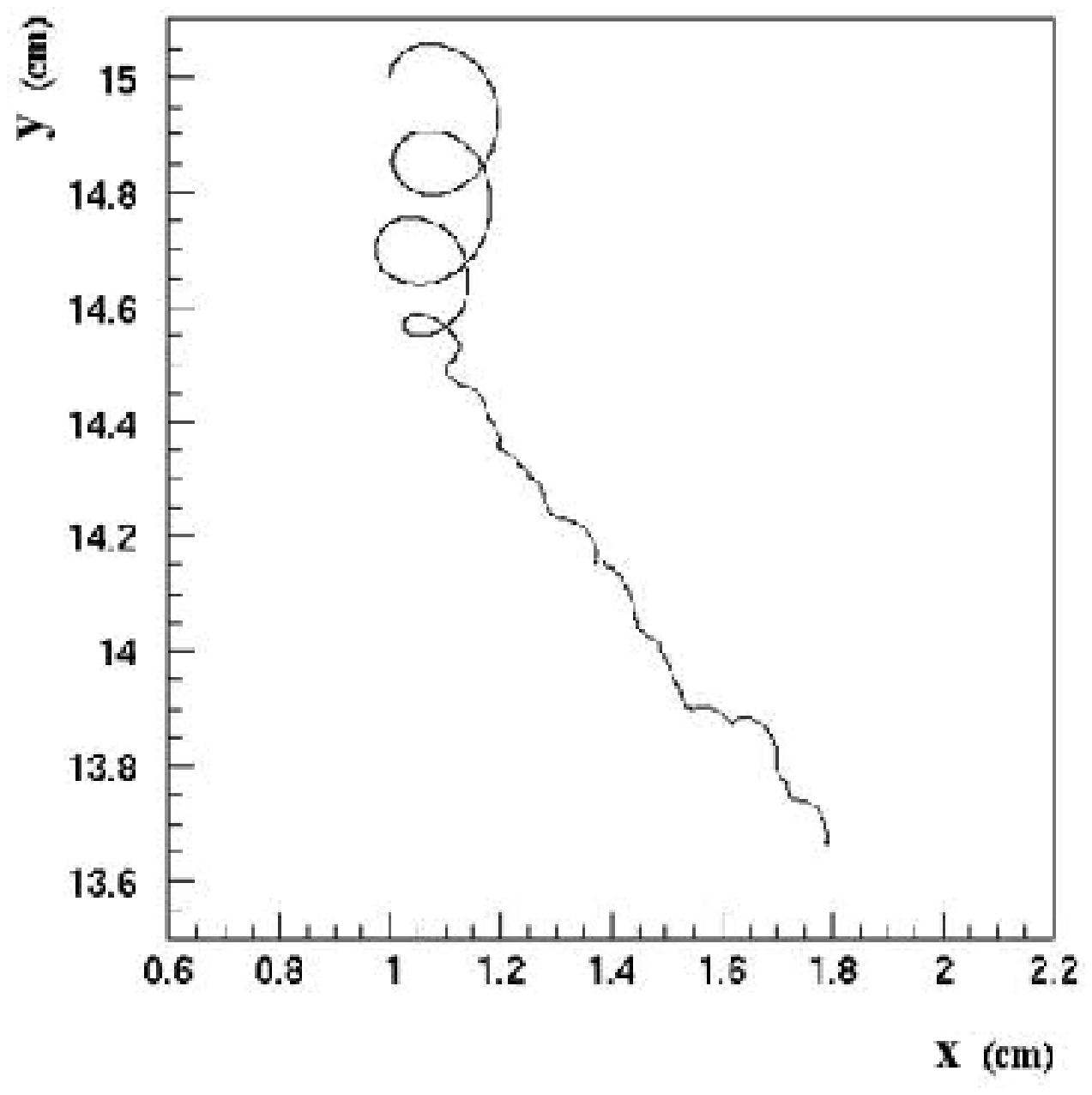,width=0.85\textwidth}
\end{minipage}
\caption{The yz (left) and xy (right) views of the final stages of the
trajectory of a single muon in the
cooling cell with
initial momenta: 
$P_x=1~{\rm MeV}, P_y=0~{\rm MeV}$ and $P_z=100~{\rm MeV}$.}
\label{fig:highmom}
\end{center}
\end{figure}

\begin{figure}
\begin{center}
\epsfig{file=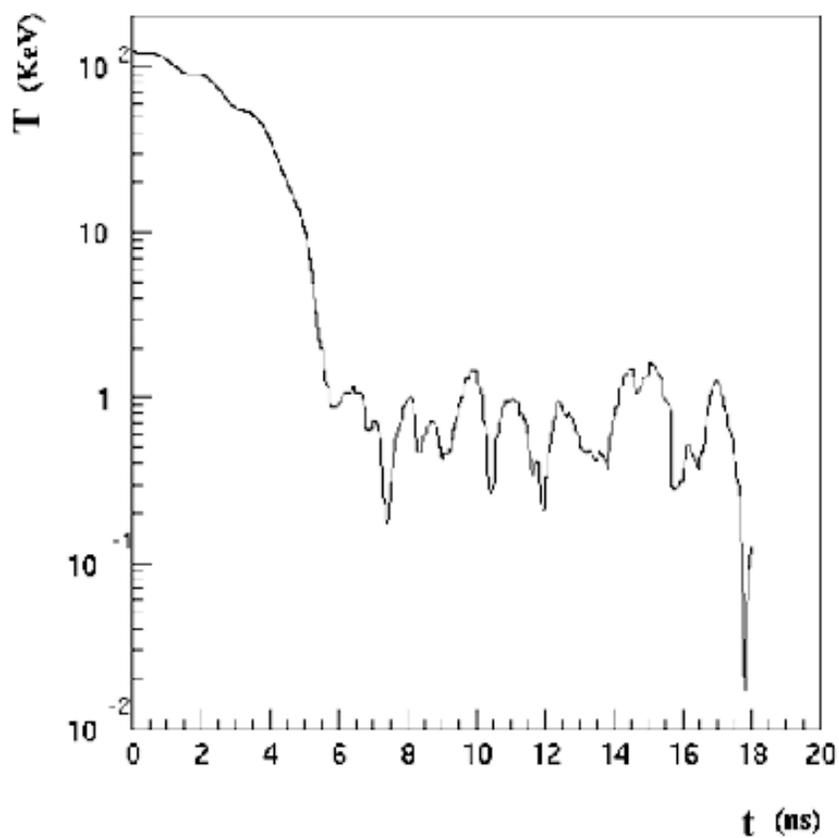,width=12cm}
\caption{The kinetic energy, $T$, as a function of time, $t$, of the
  final stages of the trajectory of a muon in the
cooling cell with initial momenta: $P_x=1~{\rm MeV/c}, P_y=0~{\rm MeV/c}$
and $P_z=100~{\rm MeV/c}$. }
\label{fig:Tkin}
\end{center}
\end{figure}

We now describe example muon trajectories in different sections of the cooling
cell.  In Fig.~\ref{fig:highmom}, the path of a muon with initial momenta:
$P_x=1~{\rm MeV/c}, P_y=0~{\rm MeV/c}$ and
$P_z=100~{\rm MeV/c}$ is shown 
in the yz and xy views.
In the yz view, the slow drift of 
the muon in the y direction resulting from the $\vec{E} \perp \vec{B}$
field is evident.  Also, the radius of the trajectory is decreasing due to
the energy loss in the gas.  Once the muon has reached the equilibrium
energy, it is extracted from the cell at a fixed Lorentz angle, which
is clearly 
seen in the x-y plane view.  
The effect of the large angle
nuclear scatters is visible.  The effect on the kinetic energy of the muon
is shown in Fig.~\ref{fig:Tkin}, where the kinetic energy is plotted as
a function of time.
The large oscillations in kinetic energy depend very
strongly on the scattering angle.  In some cases, the muon bounces off a
nucleus in a direction opposite to the accelerating force.  In this case, a
muon can come almost to rest before it is reaccelerated.  This is a
particularly dangerous situation for $\mu^-$, since the capture probability
increases rapidly at the smaller kinetic energies as shown in 
Fig.~\ref{fig:mucap}.

The muons can drift either in the positive-x or negative-x direction
depending on the sign of the electric field in the region where it
reached equilibrium energy.
Once the muon reaches the edge of the cooling cell, it is extracted through
a thin window and enters a completely evacuated region.  

\section{Acceleration and Bunching}

\subsection{Initial Reacceleration}

\begin{figure}
\begin{center}
\epsfig{file=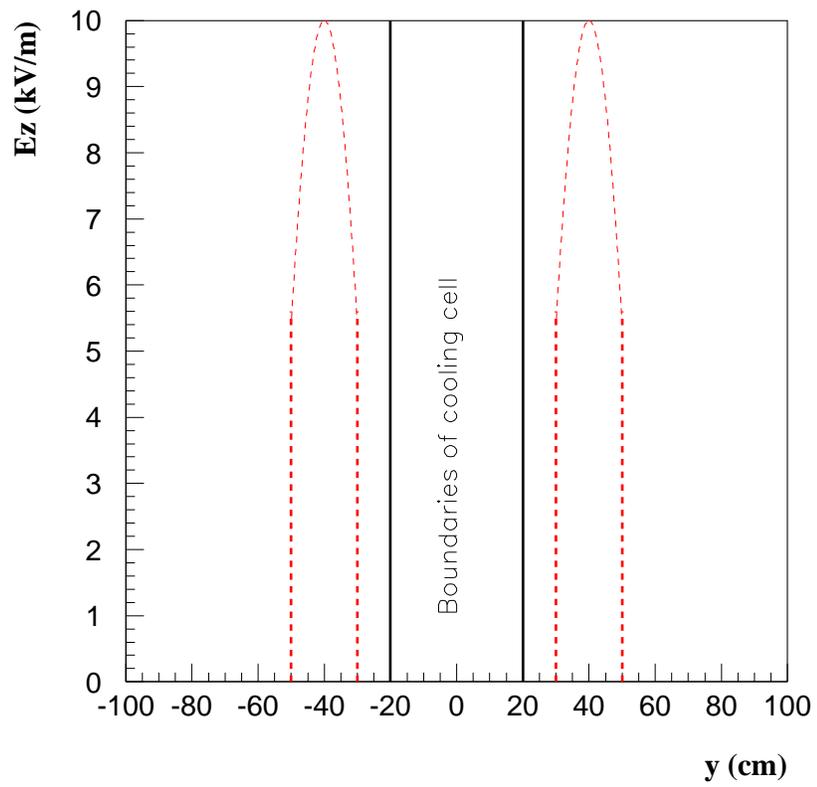,width=12cm}
\caption{Variation of E$_z$ as a function of y inside cooling cell region.}
\label{fig:ezfield}
\end{center}
\end{figure}

\begin{figure}
\begin{center}
\epsfig{file=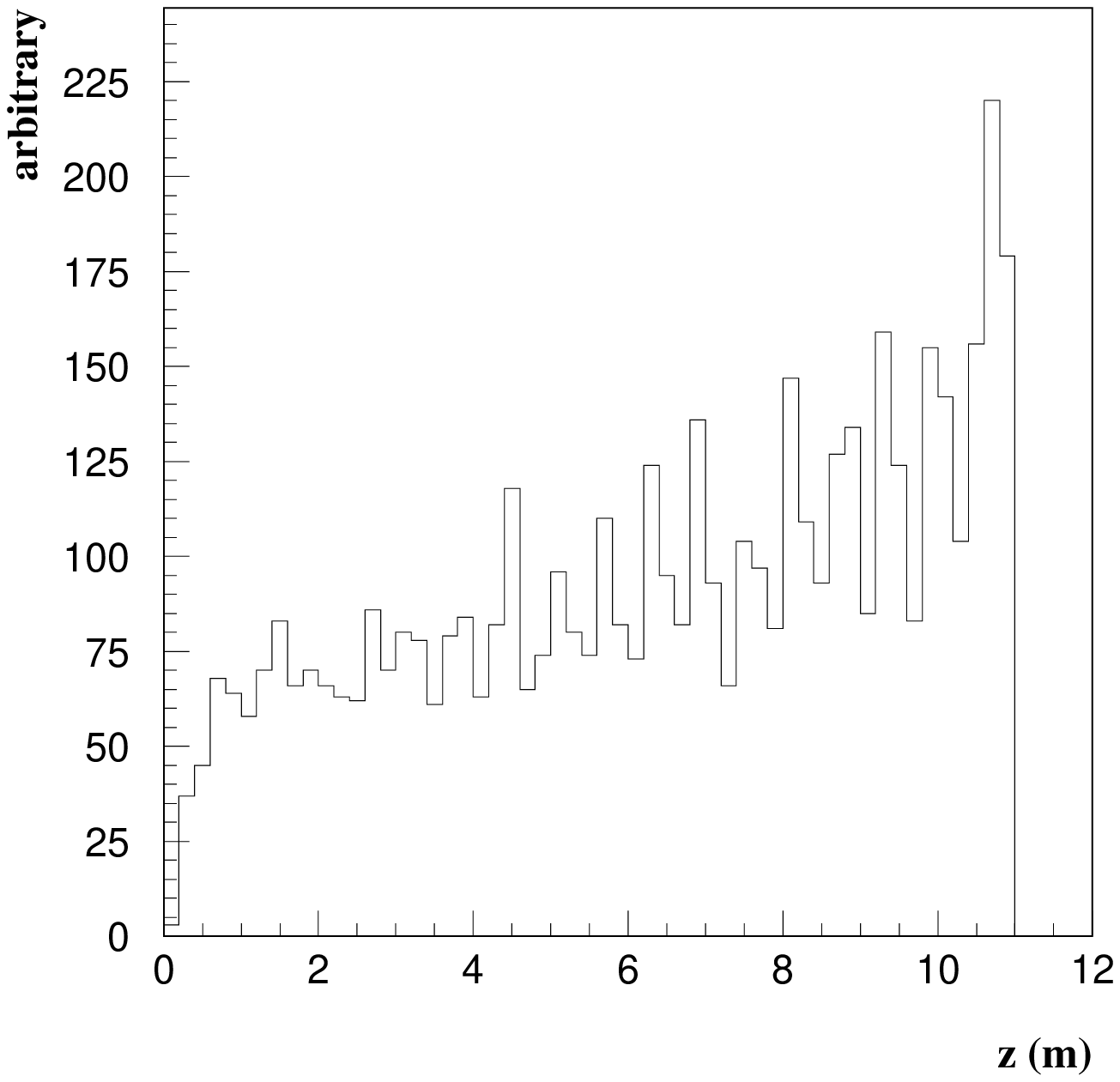,width=12cm}
\caption{The distribution of the z position for the surviving muons
extracted from the 11~m 
long cooling cell.}
\label{fig:zextract}
\end{center}
\end{figure}

Upon leaving the cooling cell, the extracted muons undergo cycloid 
motion with net
direction of motion along the positive or negative-y direction.
The parameters of the cycloid change adiabatically with the changing
electric 
field as a function of y. For $|y|>30$~cm, an electric field component in the
z-direction is also present,as shown in Fig.~\ref{fig:ezfield}, 
so that muons are then accelerated along the z-axis.  
This electric field
is relatively weak in order to avoid inducing too large a momentum
spread in the beam 
at the end of the $11$~m section. The value of the z-coordinate at
which muons are 
extracted from the gas cell is plotted in Fig.~\ref{fig:zextract}. 
The
distribution is rather flat, so that muons will have a relatively
flat kinetic energy spectrum at the end of the $11$~m cell between
$0$ and $0.11$~MeV.

\begin{figure}
\begin{center}
\epsfig{file=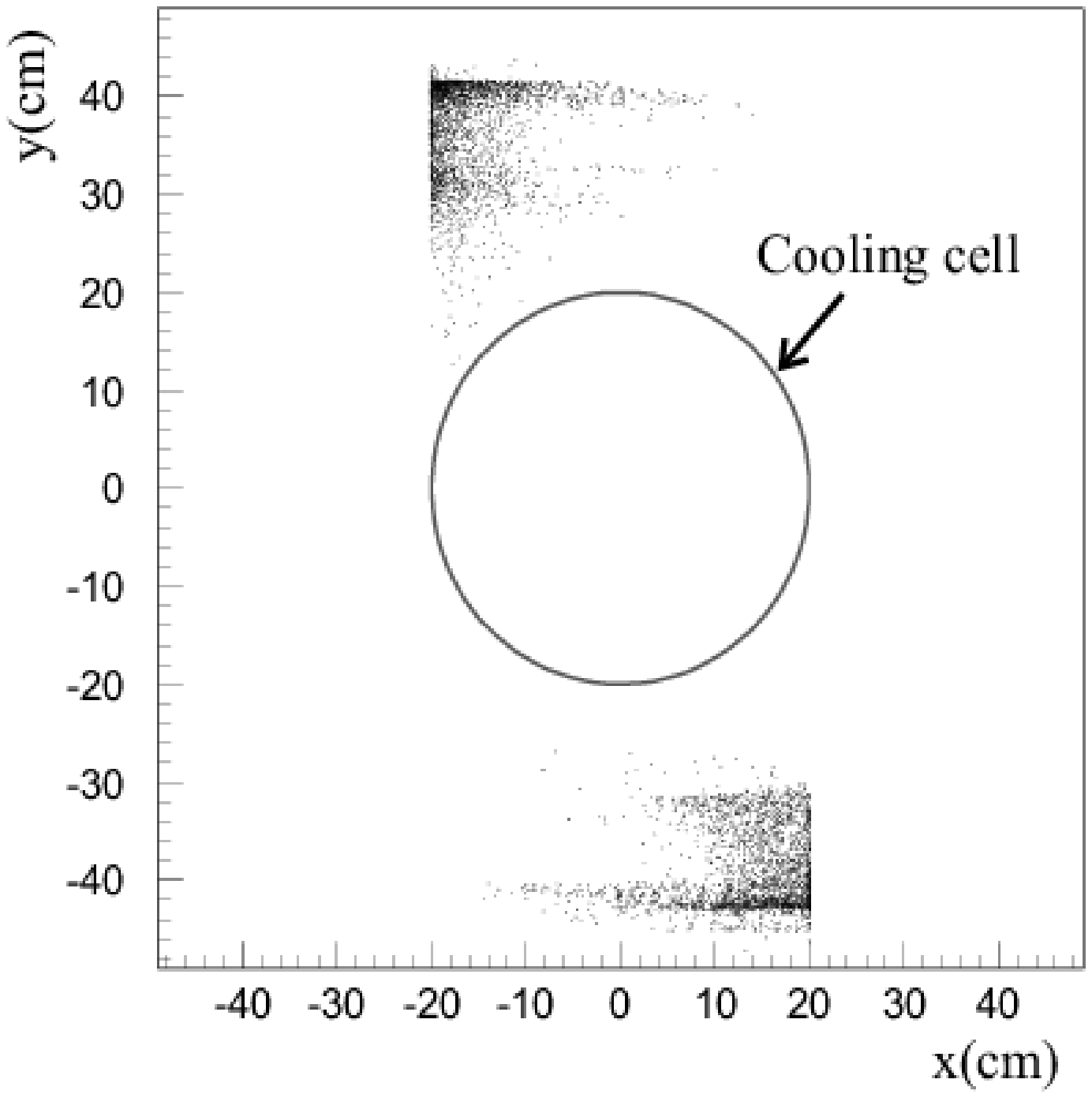,width=12cm}
\caption{The x,y distribution of muons at z=11~m after the initial
reacceleration. }
\label{fig:finalxy}
\end{center}
\end{figure}

\begin{figure}
\begin{center}
\epsfig{file=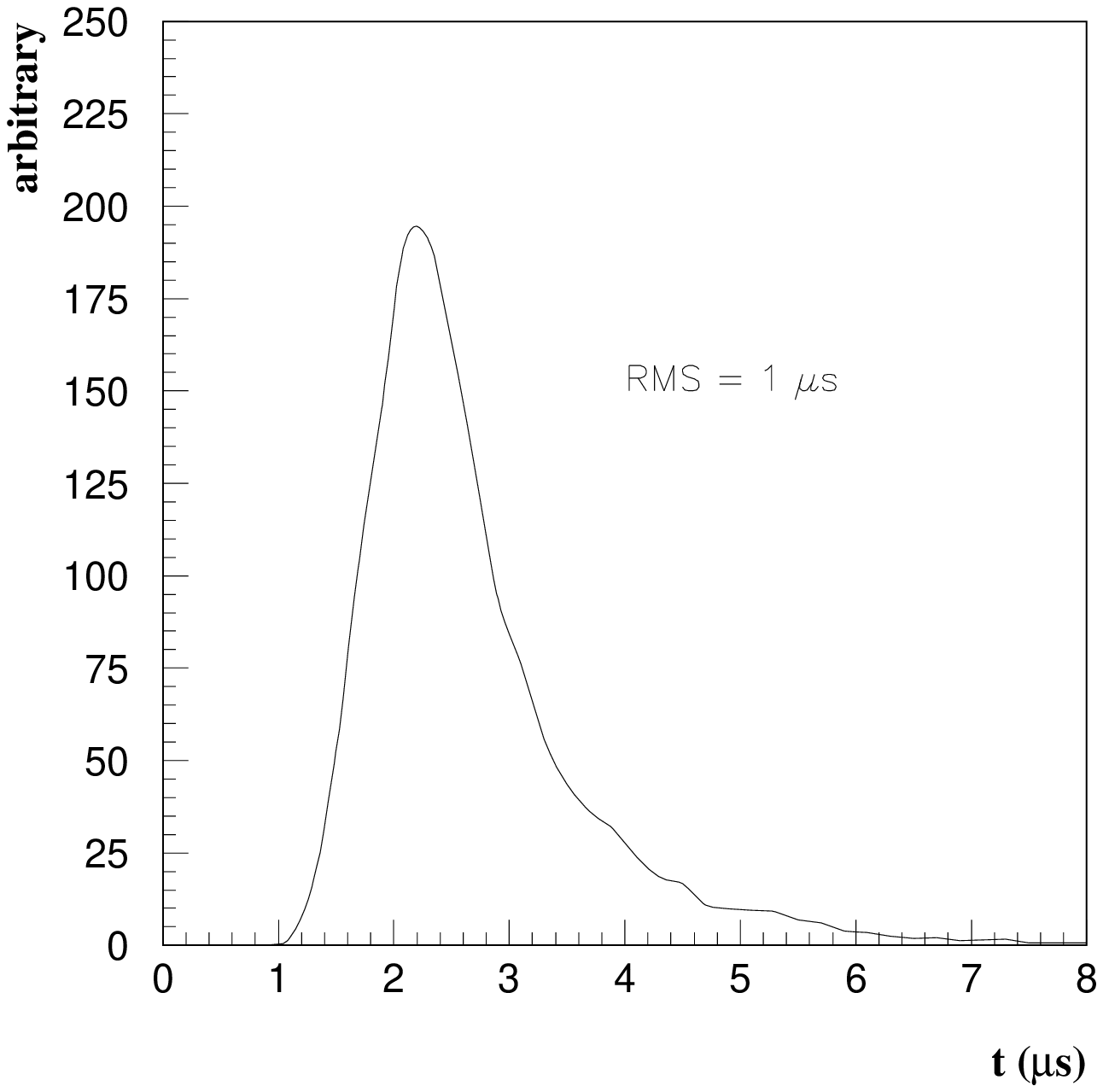,width=12cm}
\caption{The arrival time, $t$, distribution  of  muons at z=11~m
  after the initial 
reacceleration. $t=0$ is defined as the time when the proton beam hits
the target. }
\label{fig:finaltime}
\end{center}
\end{figure}

The final beam position after this
reacceleration is shown in Fig.~\ref{fig:finalxy}. The muon beam has
an arithmetic mean energy of $\sim 427$~keV and a root mean square of
$\sim 306$~keV. The time spread of the beam coming out of this initial
reacceleration is shown in Fig.~\ref{fig:finaltime} and has a root
mean square of 1~$\mu$s.

\begin{figure}[htb]
\begin{center}
\begin{minipage}[t]{0.48\textwidth}
\epsfig{file=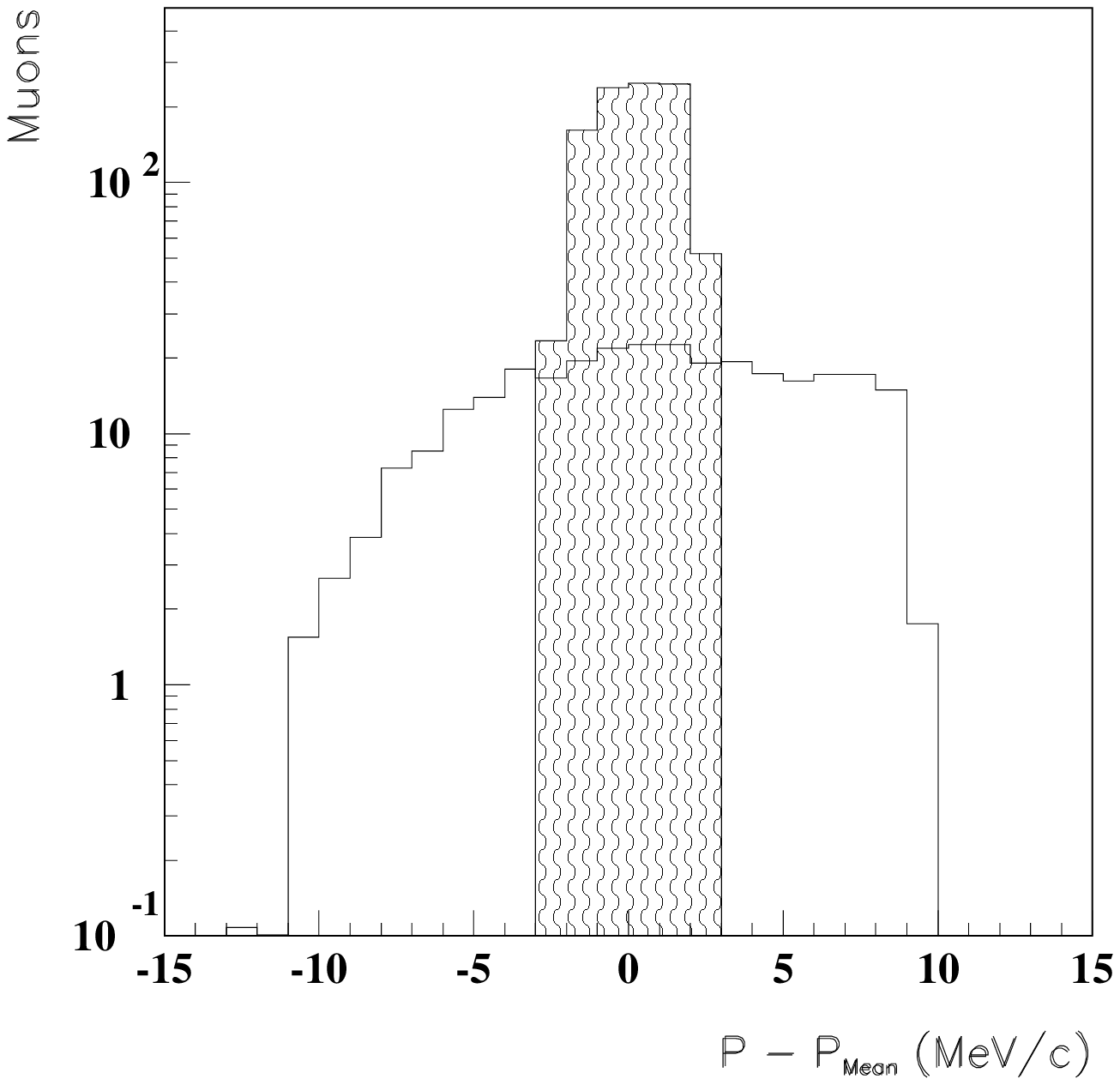,width=0.95\textwidth}
\end{minipage}
\begin{minipage}[t]{0.48\textwidth}
\epsfig{file=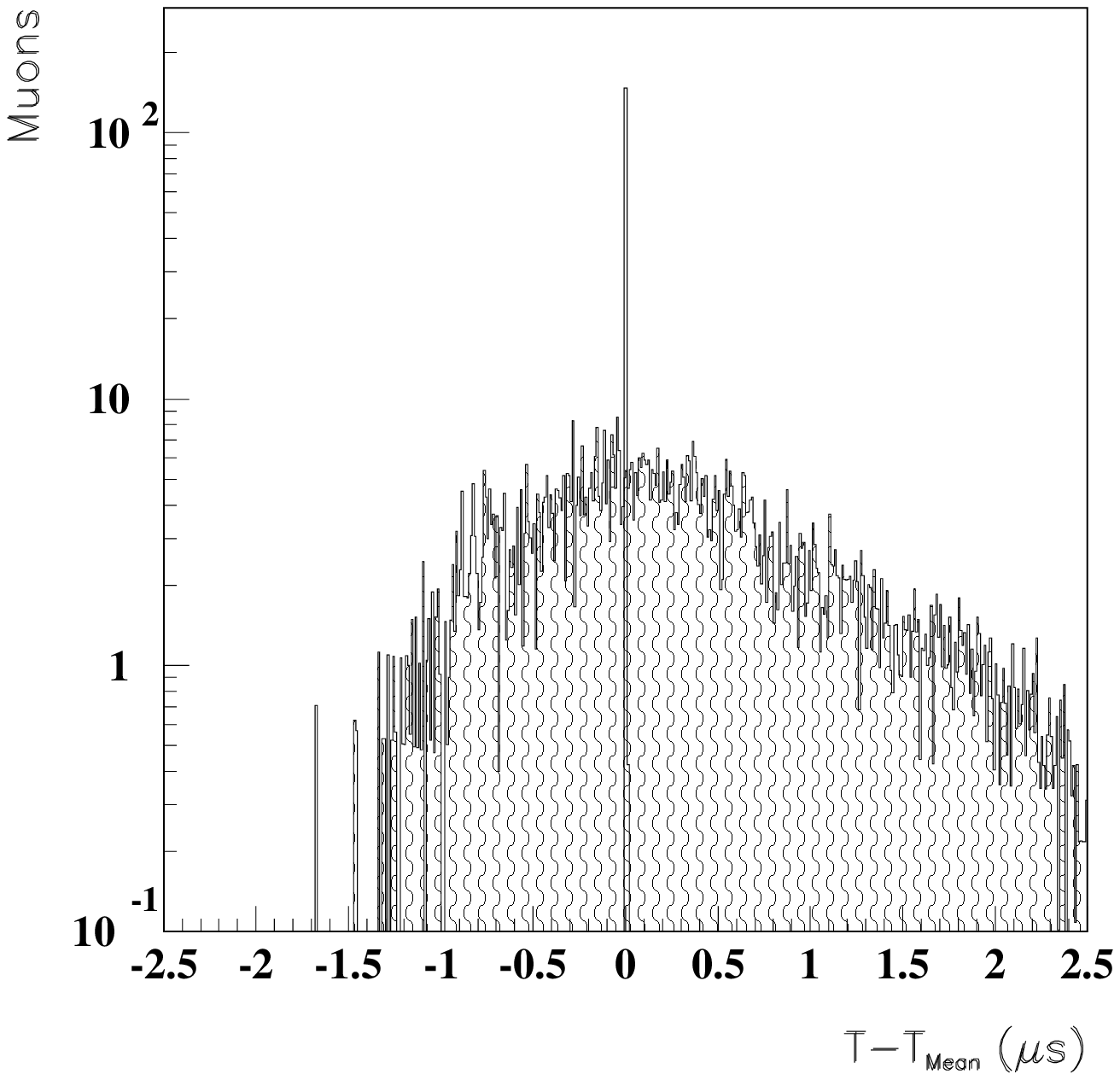,width=0.95\textwidth}
\end{minipage}
\caption{The momentum spread (left) and time spread (right) of the
muon beam before (hatched histogram) and after (histogram)
reacceleration.}
\label{fig:reaccel}
\end{center}
\end{figure}

The beam coming out of the initial reacceleration region must continue
to be
rapidly accelerated and, most importantly, the time 
spread reduced. Fig.~\ref{fig:reaccel} shows the results of a
preliminary reacceleration. The effect on the time distribution is
significant as the RMS of the time distribution goes from 1~$\mu$s to
3~ns whilst the momentum RMS spread increases from 1.2~MeV/c to
5~MeV/c at a final mean momentum of 147~MeV/c. The survival
rate for this reacceleration is 30\%. 

\subsection{Phase Space Reduction}

The six-dimensional normalized emittance  is calculated as follows:
$$\epsilon_{6D,N}=\frac{\sigma_{x}\sigma_{y}\sigma_{z}\sigma_{px}\sigma_{py}\sigma_{pz}}{(\pi~m_{\mu}c)^3},$$
where the $\sigma$'s are RMS values of the transverse and longitudinal
components of the muon beam. The emittance is evaluated at a given z
position of the beam and thus the $\sigma_z$ is calculated using
$\sigma_{\beta c t}$ at a given z. In calculating the emittance no
correlations are 
taken into account between the distributions. The total phase space
reduction is calculated by taking the ratio
($\epsilon'_{6D,N}/\epsilon^0_{6D,N}$) of the final,
$\epsilon'_{6D,N}$, and initial, $\epsilon^0_{6D,N}$, 
emittances of the muon beam. The initial emittance is calculated using
the initial distributions of those muons which are successfully
cooled. Another method would be to take the emittance calculated using
all the muons at the end of the drift. However, there is an acceptance
limitation for this channel - the channel cannot  
accept the entire beam produced by the front end of this scheme. This
is evident since the initial emittance, calculated from the
beam components given in Table~\ref{tab:muresults}, is smaller than the
emittance of the entire muon beam at the end of the drift. 
\begin{table}
\begin{center}
\begin{tabular}{||c||c||}
\hline\hline
& $\mu^+$ \\
\hline\hline
x$_0$ (cm) & 9.48  \\
y$_0$ (cm) & 2.68  \\
$\beta_0$~c~t$_0$ (cm) & 176.93 \\
px$_0$ (MeV/c) & 20.24  \\
py$_0$ (MeV/c)  & 20.39  \\
pz$_0$ (MeV/c)  & 37.91 \\
x (cm) & 4.48/5.30  \\
y (cm) & 4.76/3.85 \\
$\beta$~c~t (cm) & 1600.71/1612.02\\
px  (MeV/c) & 0.16/0.16  \\
py  (MeV/c) & 0.15/0.16  \\
pz  (MeV/c) & 1.18/1.22 7 \\
yield ($\mu$/2~GeV~p) & 0.0021\\
\hline\hline
\end{tabular}
\caption{Initial and final RMS values for the six-dimensional
ingredients of the emittance, and yields for $\mu^+$. The
final numbers for $\mu^+$ include the extraction through a 20~nm
thick Carbon window. x, y and $\beta$~c~t are the spatial
components, and px, py and pz are the momentum components of the muon
beam. Those quantities with a suffix $0$ refer to the initial values
of the muon beam and those without a suffix  refer to the final
values. Note that there are two values for each of the final
components referring to the two beamlets which come out of this scheme.
}
\label{tab:muresults}
\end{center}
\end{table}
This
indicates that any pre-cooling channel which cools the beam 
by one order of magnitude could then be followed by such a frictional
cooling channel and hence increase the yield of the entire scheme.

The results for the simulated scheme
outlined in this report are 
listed in Tables~\ref{tab:muresults} and~\ref{tab:muemitt}.
\begin{table}
\begin{center}
\begin{tabular}{||c||l|l|l||}
\hline\hline
Window thickness& & 1st Beamlet&2nd Beamlet\\
     20~nm $\mu^+$      & Initial & Final  & Final\\
\hline\hline
$\epsilon_{T,N}$ ($\pi$~m)$^2$& 9.51e-6 &4.56e-10&4.71e-10\\
$\epsilon_{L,N}$($\pi$~m)&0.20&0.057& 0.059\\
$\epsilon_{6D,N}$($\pi$~m)$^3$&1.92e-6&2.61e-11& 2.79e-11\\
\hline\hline
\end{tabular}
\caption{Normalized emittances for the initial and final muon beams
for $\mu^+$.}
\label{tab:muemitt}
\end{center}
\end{table}

The yield for
this scheme is $0.0021~\mu^+$ per 2 GeV proton. This yield does not
change greatly 
for window thicknesses up to 20~nm. As a consequence of the alternating
transverse electric field there are two beamlets emerging. The
question of the final
emittance of the total beam after recombining the beamlets is left 
open. Averaging their emittances or at
worst 
adding them together  would still result in a beam  of
sufficient emittance for luminous collisions. The total phase space
reduction, depending
on how one combines  the beamlets, ranges from  $1\cdot 10^{-5}$ to
$3\cdot10^{-5}$. This
scheme produces a cool muon beam with 
$\epsilon_{6D,N}\sim 2\cdot 10^{-11}~(\pi~{\rm m})^3$. This should be
compared to 
$1.7\cdot 10^{-10}~(\pi~{\rm m})^3$ from Table~\ref{tab:pars}, which is the
minimum emittance for luminous collisions at a Muon Collider. The
Frictional Cooling scheme outlined in this paper is an order of
magnitude better than this benchmark. The yield is a factor of 5 lower
than hoped but is compensated by the reduced emittance. 

Although earlier studies~\cite{ref:menufact02} have shown similar
promising results for 
$\mu^-$, this scheme has not been fully evaluated for $\mu^-$.

There is still room for improvement.
For example, the electric field configuration is smoothly
oscillating to make it more realistic but it is far from optimal.
Parameters, such as the length of the cell and
strength of the initial reacceleration can still be tuned together to
achieve better performance. Nonetheless, the ingredients for
successful phase space reduction of muon beams are there.

\section{Issues for Future Studies}

This paper is meant to outline a possible phase space reduction scheme
for a muon beam intended for a Muon Collider. Amongst the 
improvements and refinements of the scheme,
there are several issues which require further thought and eventually
experimental proof of principle.

\subsection{Windows}
\label{sec:windows}

The windows must be
thin, gas tight and, eventually, made larger in area than currently
available. 

A  Frictional Cooling demonstration experiment using
protons~\cite{ref:me}  was 
performed at Nevis Laboratories. The experiment used thin windows
which were 20~nm Carbon on a thin Nickel grid, as quoted from the
manufacturer. However, the data indicated an effective thickness which
was more than an order of magnitude larger than what was expected,
resulting in 
all protons which achieved the equilibrium being stopped in the exit
window. This complication further emphasizes the importance of the
window issue.

\subsection{Muon Capture}

The capture cross section of the $\mu^-$ by He or H$_2$ gas, at
energies below 1~keV, 
is experimentally unknown. 
Theoretical calculations were incorporated in the simulation programs
but it remains an experimental issue for this scheme. We anticipate
performing an experiment to directly measure this capture cross section.

\subsection{Electrical breakdown}

Most accelerating structures in particle accelerators (RF cavities
etc.) operate in as good a vacuum as possible to avoid electrical
breakdown.  Conversely, breakdown can also be suppressed by using
dense materials between electrodes. In this case, the mean free path
between collisions for free ions is so small that the ions cannot
accelerate to high enough energies to create an avalanche. The Paschen
law~\cite{ref:paschen} defines the breakdown voltage in gases as 
a function of pressure and distance. For 1~atm of Helium, the electric
field achievable before breakdown is less than $\sim 0.5$~MV/m. This
is an order of magnitude less than what has been considered in the
Frictional Cooling cell. However, in a crossed ${\vec E}$ and ${\vec
B}$ field, a charged particle 
experiences cycloid motion resulting in a maximum kinetic energy
defined by $T_{max}=
2m(|\vec{E}|/|\vec{B}|)^2$, where $m$ is the mass of the charged particle. Hence, the
maximum kinetic 
energy that can be achieved by an ionized electron for  $|\vec{B}|=5$~T and
$|\vec{E}|=5$~MV/m is 11.4~eV. This is below
the ionization energy of both Helium and Hydrogen (25.4~eV and 13.6~eV
respectively) and hence an ionized electron would not create an
electrical breakdown in the gas. It should therefore be possible to
suppress multiplication in our cooling cell.

\subsection{Plasma Formation and Charge screening}
\label{sec:plasma}

The crossed $\vec{E}$ and $\vec{B}$ fields may allow the bounds from
the Paschen Curves to be avoided but there is still a large amount of
energy being deposited into the gas cell which will create a large
number of free ions. The danger is not that avalanche and breakdown
will occur but rather that the
separation of the free charges may screen or at worst cancel the
imposed electric field. 

To meet the final luminosity requirements, 
of order $5\cdot 10^{12}$ muons per pulse will have to be stopped in the 
gas cell, with a mean initial 
kinetic energy of $10$~MeV.
Taking an ionization
energy of $25$~eV for Helium, we find $2 \cdot 10^{18}$
ionized electrons.  
This is a large number of
electrons and ions, which will drift apart in the crossed electric and
magnetic field, tending to screen the field. The speed at which this occurs, 
and the effect on the extraction of the muons, will clearly need careful
evaluation.

\section{Summary and Conclusions}

A Muon Collider would be extremely valuable to the field of High
Energy Physics, leading to a new era of precision investigation
whilst opening up the high energy frontier for discovery. Any Muon
Collider will require an efficient and effective cooling 
scheme for muons. 

The scheme described here is based on the concept of
Frictional Cooling. To that end simulations have been performed to
evaluate Frictional Cooling as well as the components required
upstream of the cooling module, such as the proton driver. We propose
a 2~GeV proton beam impinging of a Copper target as the front end. By
capturing the low energy pion cloud transverse to the target,
relatively equal yields of both $\pi^+$ and $\pi^-$ are produced. This
allows the development of a symmetric machine. The final
six-dimensional emittance coming out of the cooling section is 
$2\cdot 10^{-11}~(\pi~{\rm m})^3$, which is better than the target
emittance in various parameters sets of potential Muon Colliders. The
yield for this scheme, as it was simulated for this study, is
$0.0021~\mu^+/2$~GeV proton. This is somewhat low but there is
potential for improvement.
A rapid reacceleration of the muon beam has also been developed and
takes the cool muon beam to a mean momentum of 147~MeV/c with a
survival probability of 30\%. The simulation code has been
experimentally supported by data from a Frictional Cooling  experiment
with protons~\cite{ref:me} and more experiments are planned to further
test the Monte Carlo. 

The results outlined in this paper show that a
muon beam of sufficiently small emittance for luminous collisions can
be produced. The steps from production, cooling and reacceleration
have been addressed and a series of critical issues have also been
outlined for this scheme to be further developed. At this stage the
results are encouraging and Frictional Cooling continues to be an
exciting alternative potential for the successful phase space
reduction of muon beams intended for a Muon Collider. 

\section{Acknowledgments}

This work was funded through an NSF grant number  NSF PHY01-04619,
Subaward Number  
39517-6653. We were fortunate enough to have summer students in 2001
and 2002 who participated in this work. They include Emily
Alden, Christos Georgiou, Daniel
Greenwald, Laura Newburgh, Yujin Ning, William Serber and Inna Shpiro.
Nevis Laboratories acted as the host lab for the bulk of the work
described in this paper. The authors would like to thank Nevis
Laboratories and its staff for facilitating this research.

\end{document}